%% file: IntroISE.tex
\documentclass[12pt,a4paper]{article}
\usepackage{amsmath}
\usepackage{amssymb}
\usepackage[usenames, dvipsnames]{color}
\usepackage{mathtools}
\usepackage{verbatim}
\usepackage{amsfonts}
\usepackage{mathrsfs}
\usepackage{graphicx}
\usepackage[caption=false]{subfig}
\usepackage{enumitem}
\usepackage[normalem]{ulem}
\usepackage[percent]{overpic}
\usepackage{bm}
\usepackage{bbm}
\usepackage{rotating}
\usepackage{hyperref}
\usepackage{cancel}
\usepackage{verbatim}
\usepackage{graphicx}
\usepackage{tensor}
\usepackage{wrapfig}
\usepackage{booktabs}
\usepackage{natbib}
\usepackage{setspace}
\usepackage{relsize}
\usepackage{authblk}
\usepackage[hang,flushmargin]{footmisc}

\usepackage{geometry}
\usepackage{braket}
\newcommand{\ii}{\mathrm{i}}
\newcommand{\openone}{\mathbbm{1}}
\renewcommand{\d}{\mathrm{d}}

\newcommand{\ToDo}[1]{{\color{red}\bf [To Do: #1]}}
\newcommand{\surint}{\xrightarrow[\text{int. as}]{\text{survives}}}

\begin{document}
\title{Introducing the ISE Methodology:\linebreak
A Powerful New Tool for Topological Redescription}
\author{Daniel Grimmer}\footnotetext{\ \ Email: daniel.grimmer@philosophy.ox.ac.uk}
\affil{\small University of Oxford, Reuben College}
\date{}

\maketitle
\begin{abstract}
This paper introduces a powerful new tool for topological redescription, the ISE Methodology. These tools allow us to remove and replace a theory's topological underpinnings just as easily as we can switch between different coordinate systems. Aspirationally, these novel topological redescription techniques can be used to provide new support for a roughly Kantian view of space and time; Rather than corresponding to any fundamental substances or relations, we can see the spacetime manifolds which appear in our theories as merely being an aspect of how we \textit{represent} the world. This view of spacetime topology parallels the dynamic-first view of geometry as well as a Humean view of laws; The spacetime manifolds which feature in our best theories reflect nothing metaphysically substantial in the world beyond them it being one particularly nice way (among others) of codifying the dynamical behavior of matter.

A parallel publication (namely,~\cite{ISEEquiv}) will explicitly characterize the power and scope of the topological redescription techniques offered to us by the ISE Methodology. The modest goal of this paper is simply to introduce the ISE Methodology by applying it to two example theories. Firstly, to familiarize ourselves with these techniques, I will show how they can be used to redescribe a spacetime theory via a Fourier transform. Secondly, I will show how the exact same techniques can be used to redescribe a lattice theory (i.e., a theory set on a discrete spacetime, $\mathcal{M}_\text{old}\cong\mathbb{R}\times\mathbb{Z}$) as existing on a continuous spacetime manifold, $\mathcal{M}_\text{new}\cong\mathbb{R}\times\mathbb{R}$.
\end{abstract}


\begingroup
\let\clearpage\relax
\input{ChapAndApp-Arxiv/0_Introduction}
\endgroup


\bibliographystyle{dcu}
\bibliography{references}


\appendix

\input{ChapAndApp-Arxiv/A1_Homo}

\end{document}

%% file: ChapAndApp-Arxiv/0_Introduction.tex
\section{Introduction}\label{SecIntro}
Since at least Newton, our best physical theories have typically featured some piece of smooth topological background structure (namely, a spacetime manifold) either implicitly or explicitly. There are, of course, several ways in which one can interpret these spacetime manifolds. Three broad classes of interpretation are as follows: substantivalism, relationalism, and a third view which I shall here call \textit{spacetime codificationism}. Views of roughly these kinds can be traced back to Newton, Leibniz, and Kant respectively.

A substantivalist views spacetime as a collection of topologically interconnected spacetime points (perhaps understood anti-haecceitistically). By contrast, relationalists would distance themselves from talk of spacetime points, instead opting to view the spacetime in terms of spatiotemporal relation. The third view, spacetime codificationism, holds that rather than corresponding to some fundamental substances or relations in the world, we ought to think of the spacetime manifolds which routinely appear in our best scientific theories as merely being an aspect of how we \textit{represent} the world. This view of spacetime topology parallels a broadly dynamic-first view of geometry as well as a roughly Humean view of laws; The spacetime manifolds which feature in our best theories reflect nothing metaphysically substantial in the world beyond them it being one particularly nice way (among others) of codifying the dynamical behavior of matter.

However one understands the spacetime manifold, the fact remains that our well-established physical theories near universally feature a designated spacetime arena within which the world's events happen (or, at least, are described to happen). Given the seeming indispensability of the spacetime manifold in physics, adopting a straightforward realist attitude naturally pushes one towards some form of spacetime substantivalism. The relationalist can, of course, push back in a variety of ways. Indeed, much has been written in the philosophy of spacetime literature about the substantivalism vs relationalism debate. The goal of this paper, however, is to introduce some new mathematical techniques which can be used in support of a roughly Kantian position: spacetime codificationism.

My target shall be the assumed indispensability of the spacetime manifold in our physical theories. To this end, this paper will introduce some powerful new tools for topological redescription (namely, the ISE Methodology). As a piece of mathematical machinery, the goal of the ISE Methodology is to offer us as many candidate spacetimes as possible, whatever we may think of them. In particular, I claim that by using the ISE Methodology one can remove and replace a theory's topological underpinnings just as easily as one can switch between different coordinate systems.

Let us suppose for the moment that the ISE Methodology meets its technical aims (as~\cite{ISEEquiv} argues it does). Conceivably, one could adopt a wide range of different philosophical responses to this sudden over-abundance of alternative spacetime descriptions. My recommendation would be spacetime codificationism, but what is this view exactly? Fortunately, a rough analog of this view is already well known:\footnote{As mentioned above, the spacetime codification view which I am advocating here is also intended to be analogous to the dynamics-first view of geometry put forward by~\cite{RBrown2005}. Indeed, this project can be seen as a response to \cite{Norton2008} by extending the dynamics-first view of geometry to a dynamics-first view of topology. See~\cite{ISEEquiv} for further discussion.} namely a broadly Humean view of the laws of nature. On such a view, the laws of nature which appear in our theories reflect nothing metaphysically substantial in the world beyond them being one particularly nice way (among others) of codifying the dynamical behavior of matter. One might then pick out the theory's fundamental laws of nature as being the its best axiomization (in some yet-to-be-specified sense of best). Note the role that our capacity for logical re-axiomization plays in supporting this view: Before we can pick out the best axiomization, we first need access to the complete range of possible axiomizations. The ISE Methodology aims to play an analogous role in a topological context.

It is beyond the scope of this paper to demonstrate that the ISE Methodology meets its philosophical aims (i.e., to support a broadly Kantian view of spacetime which is analogous to a roughly Humean view of laws). Indeed, it is also beyond the scope of this paper to prove that the ISE Methodology meets its technical aims (i.e., to give us a capacity for topological redescription on par with our existing capacity for coordinate redescription and re-axiomization). A strong argument for this second point can be found in \cite{ISEEquiv}. The modest goal of this paper is merely to get the ISE Methodology on the table, so to speak. Namely, my goal is simply to introduce these topological redescription techniques by applying them to two example theories. 

Sec.~\ref{SecPreview} will provide a preview of the ISE Methodology (whose three steps are Internalize, Search, and Externalize, hence the initials). Secs.~\ref{SecSurInt} and~\ref{SecDemoExt} will then introduce the internalization and externalization processes respectively by applying them to several example theories. For the sake of pedagogy, the first example application will be rather tame: It ultimately amounts to taking a Fourier transform. The real power of these techniques however is in their scope and generality. In Sec.~\ref{SecDemoISE} I will use the exact same topological redescription techniques to do something more interesting. I will there redescribe a lattice theory (i.e., a theory set on a discrete spacetime, $\mathcal{M}_\text{old}\cong\mathbb{R}\times\mathbb{Z}$) as existing on a continuous spacetime manifold, $\mathcal{M}_\text{new}\cong\mathbb{R}\times\mathbb{R}$. Finally, Sec.~\ref{SecConclusion} concludes.

\section{Previewing the ISE Methodology: Internalize, Search, Externalize}\label{SecPreview}
The first step of the ISE Methodology is internalization. Internalization aims to divorce a theory's dynamical and kinematical content from any assumed topological background structure (i.e., the spacetime manifold, $\mathcal{M}_\text{old}$). In spirit, this first step parallels the algebraic approach to spacetime. A key difference, however, is that the ISE Methodology can be applied to spacetime theories with no algebraic structure what-so-ever (see Sec.~\ref{SecSurInt}). Indeed, the ISE Methodology does not make use of any algebraic structure in the construction of new spacetime settings (nor in the re-construction of old spacetime settings). Instead, new spacetime settings are to be built solely from our ability to smoothly vary the theory's states (see Sec.~\ref{SecDemoExt}).

This brings us to the second step of the ISE Methodology, searching for the building blocks of a new spacetime setting. I will call these \textit{pre-spacetime translation operations} (PSTOs). PSTOs are a certain kind of smooth transformation among the theory's states. Roughly, they are a pair of Lie groups which act smoothly on the theory's states and which are, in a sense, structurally indistinguishable from spacetime translations. (A technical definition will be given in Sec.~\ref{SecDemoExt}.) For example, consider translations not in spacetime, but rather in Fourier space. This and other examples of PSTOs are sketched below in Fig.~\ref{FigPSTOs}. Before discussing these, however, allow me to first quickly overview how these PSTOs can be used to create a new spacetime setting for the theory in question.

The third step of the ISE Methodology is externalization. Having picked out some PSTOs, externalization builds a new spacetime setting from them simply by taking them seriously as spacetime translations. By construction, one's chosen pre-spacetime translation operations will end up becoming honest-to-goodness spacetime translations in the new theory. In the Fourier example, the new theory's spacetime is effectively a copy of the old theory's Fourier space. Correspondingly, the new theory's states and dynamics are naturally related to those of the old theory by a Fourier transform (see Sec.~\ref{SecDemoExt}).


Let us next see some examples of PSTOs. As a first example, consider smoothly shifting the theory's states with respect to the theory's old spacetime manifold. To make things concrete, consider a theory about a complex scalar field, $\varphi_\text{old}:\mathcal{M}_\text{old}\to \mathbb{C}$, with $\mathcal{M}_\text{old}\cong \mathbb{R}^2$. In some fixed global coordinate system, $(t,x)$, we might smoothly shift the theory's states as $\varphi_\text{old}(t,x)\mapsto \varphi_\text{old}(t-\Delta t,x-\Delta x)$ for any $\Delta t, \Delta x\in\mathbb{R}$. The action of this Lie group on $\varphi_\text{old}(t,x)$ is shown in the first column of Fig.~\ref{FigPSTOs}. In general, the old theory's spacetime translations trivially qualify as PSTOs. As \cite{ISEEquiv} proves, by externalizing these sorts of PSTOs one can easily return back to the theory's old spacetime framing. This, however, is not compulsory. We can choose to externalize different PSTOs.

\begin{figure}[p!]
\centering 
\includegraphics[height=0.11\textheight]{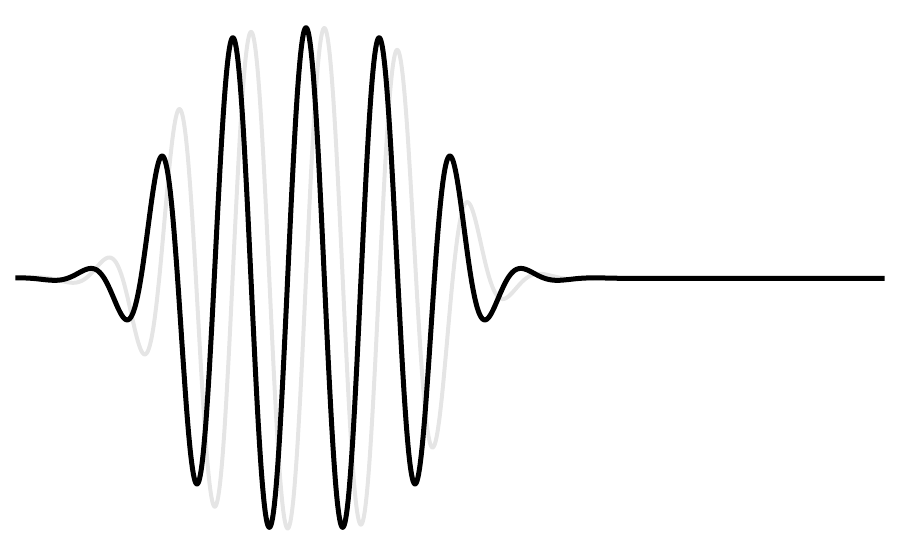}
\includegraphics[height=0.11\textheight]{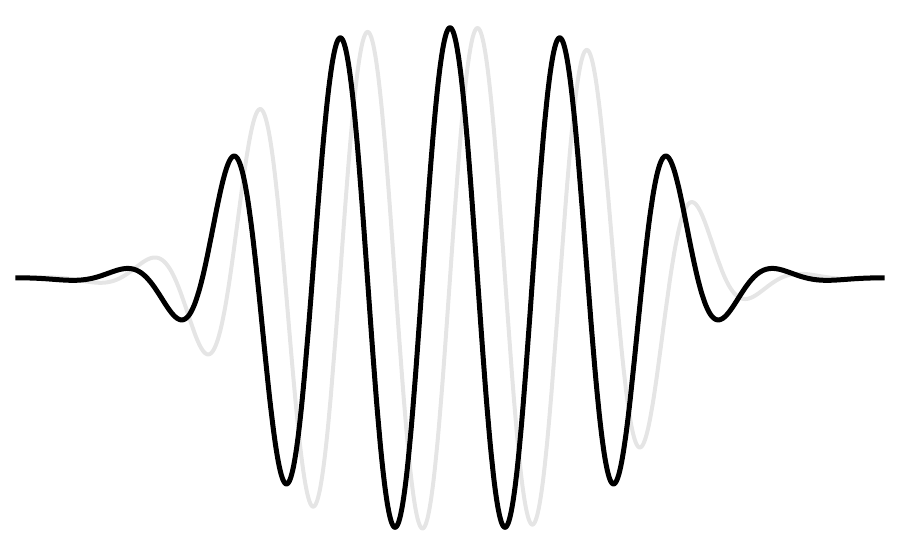}
\includegraphics[height=0.11\textheight]{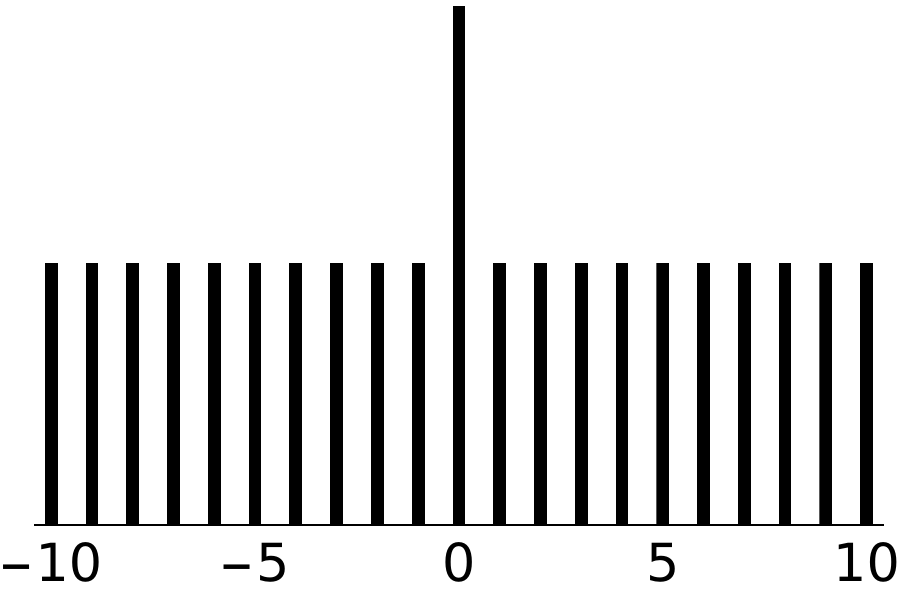}\\
\includegraphics[height=0.11\textheight]{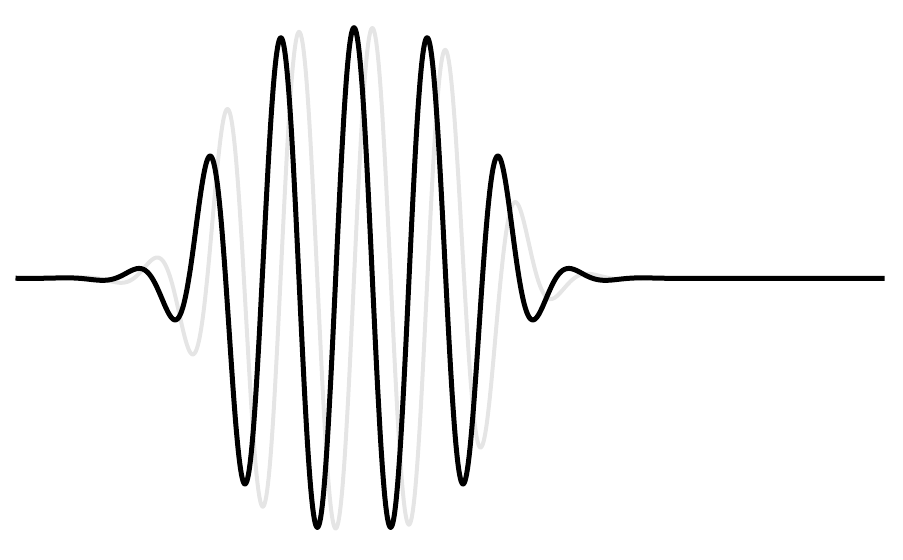}
\includegraphics[height=0.11\textheight]{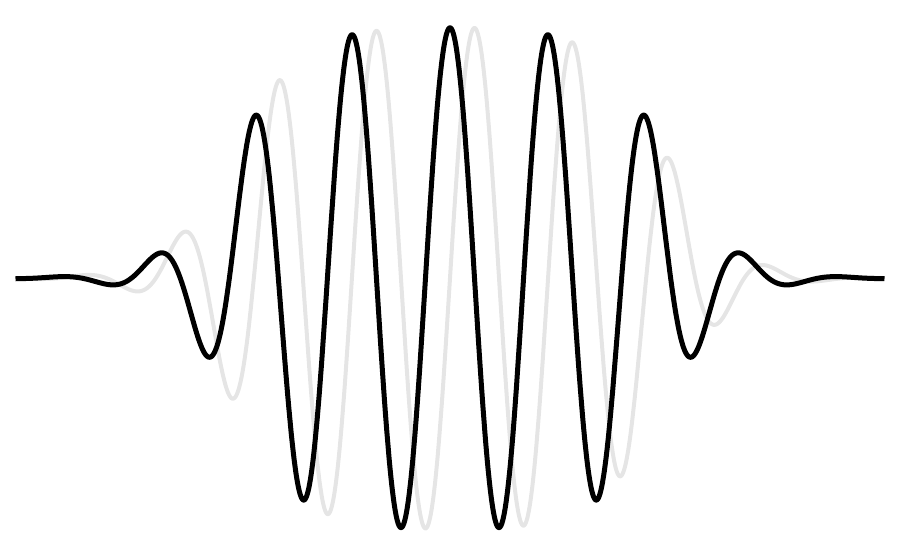}
\includegraphics[height=0.11\textheight]{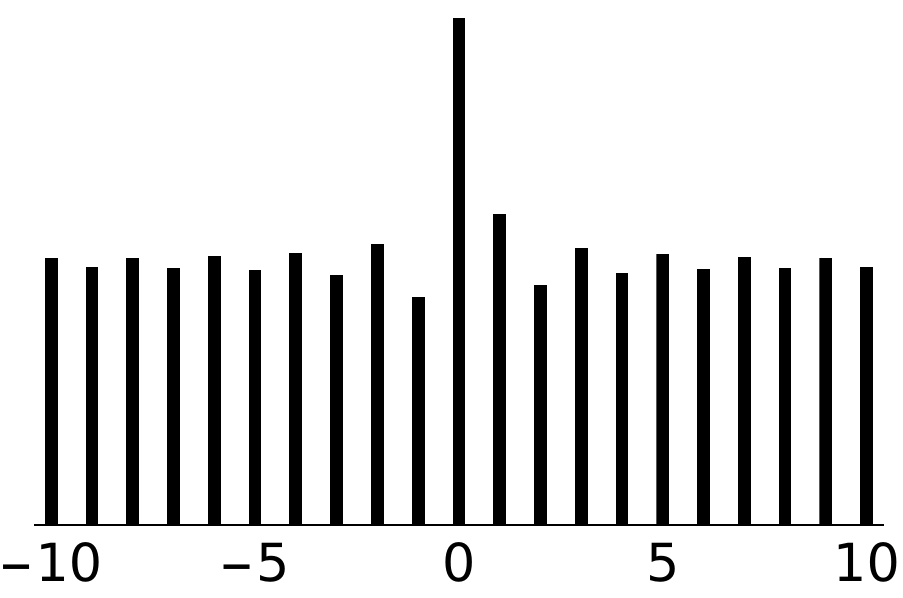}\\
\includegraphics[height=0.11\textheight]{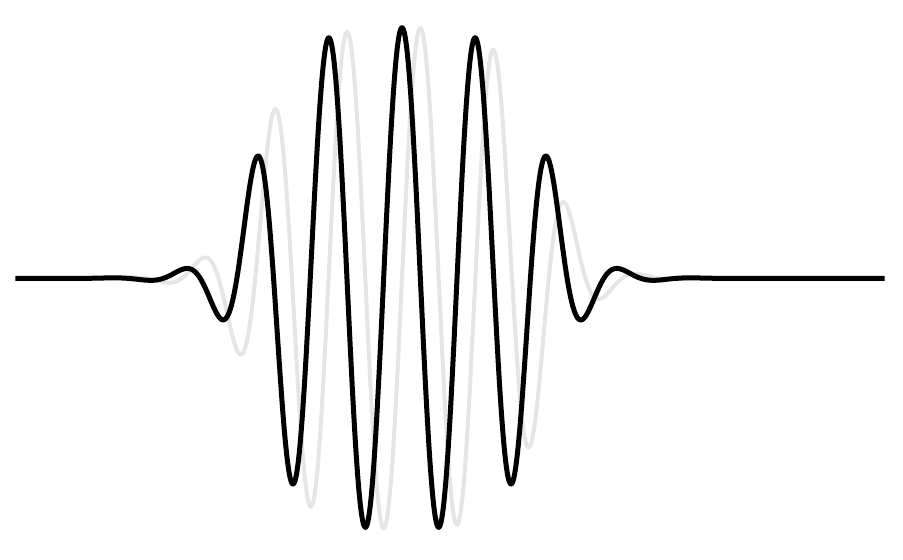}
\includegraphics[height=0.11\textheight]{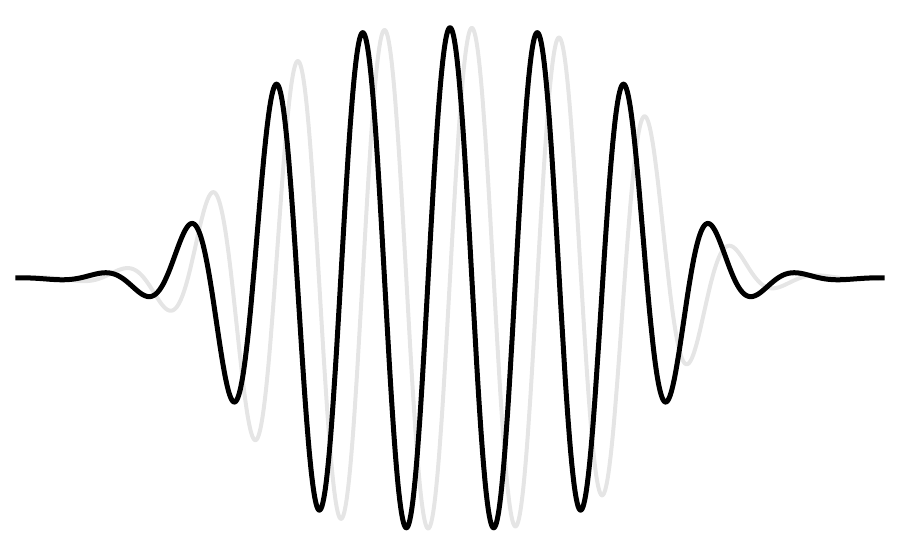}
\includegraphics[height=0.11\textheight]{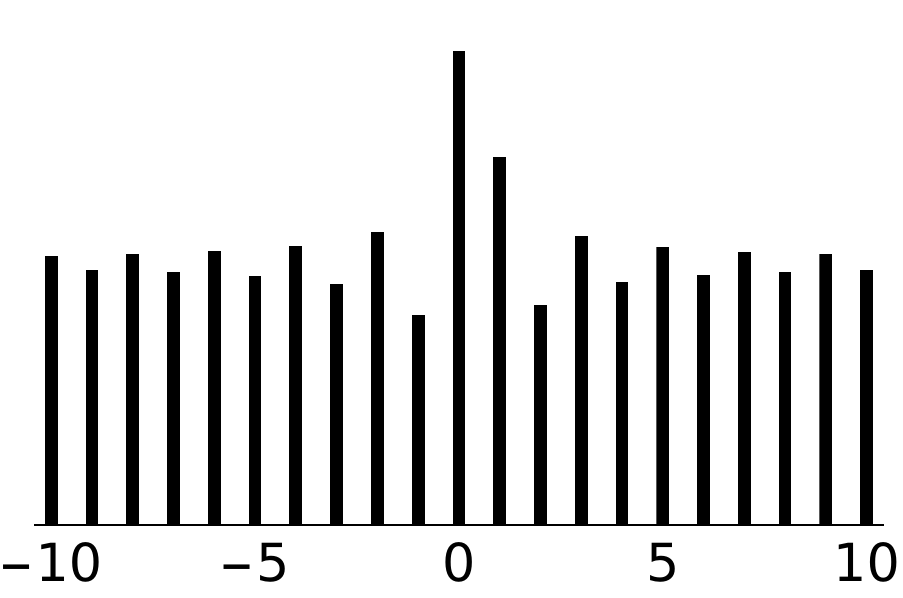}\\
\includegraphics[height=0.11\textheight]{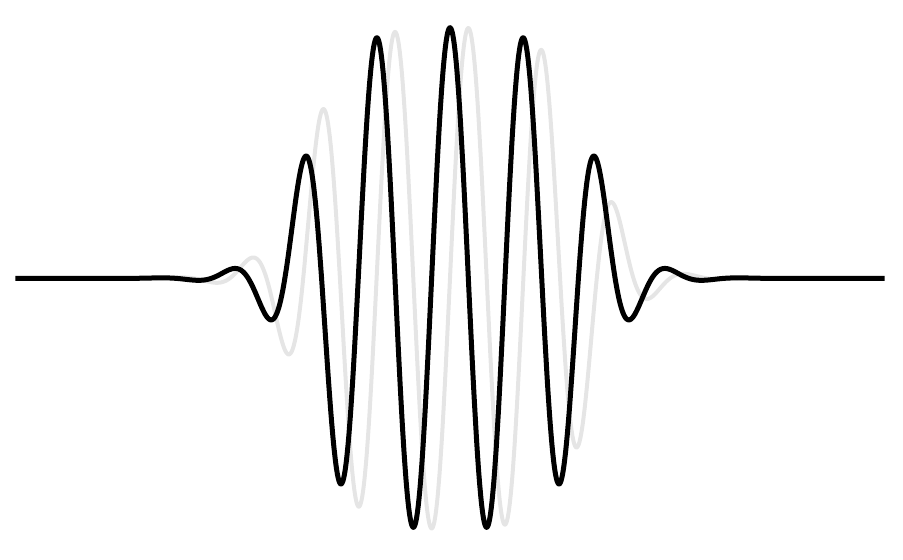}
\includegraphics[height=0.11\textheight]{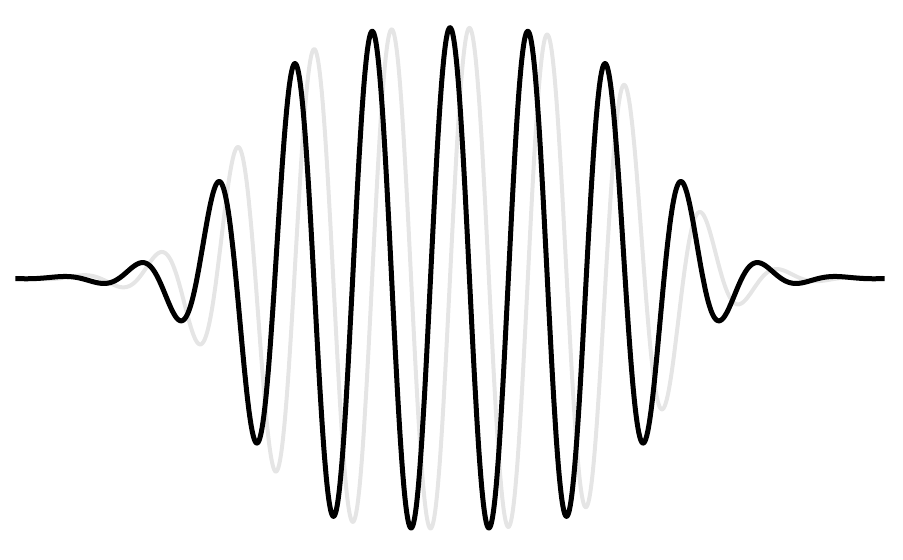}
\includegraphics[height=0.11\textheight]{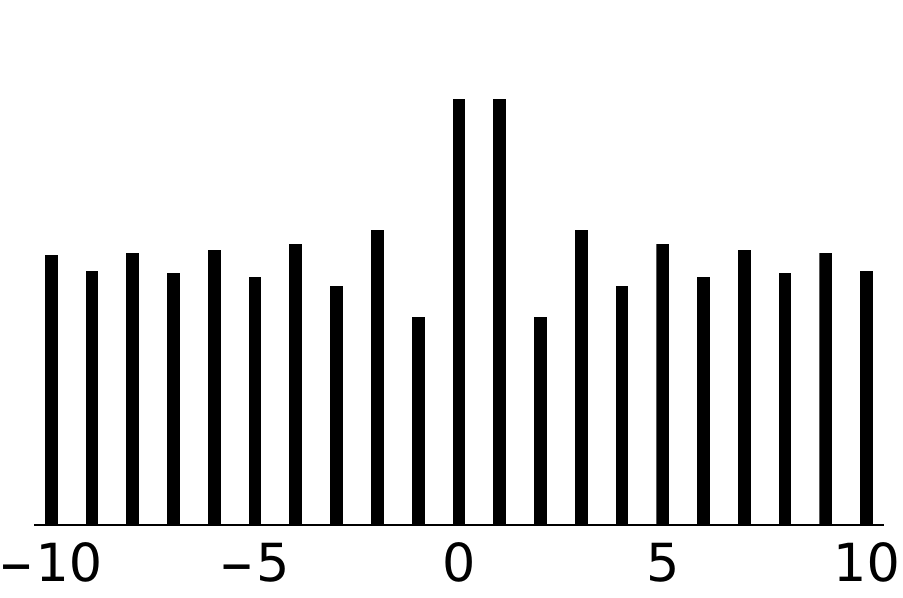}\\
\includegraphics[height=0.11\textheight]{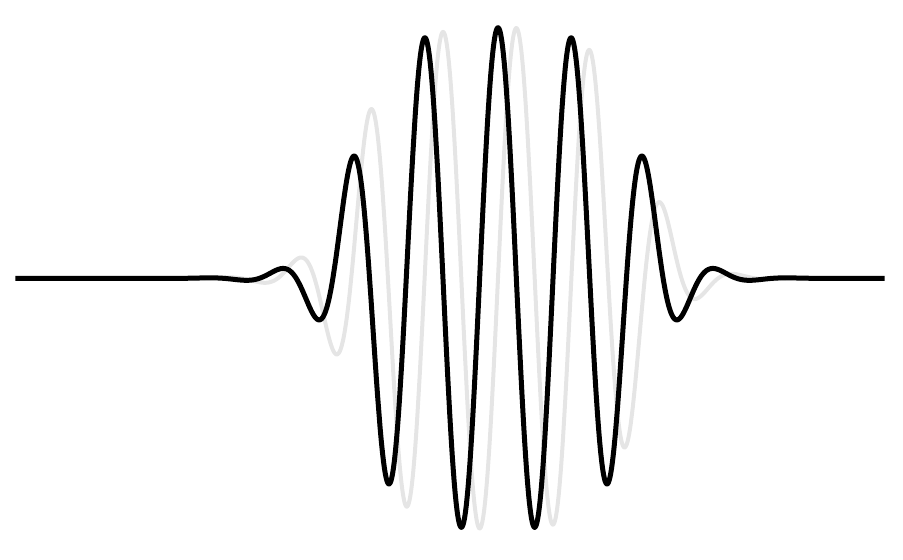}
\includegraphics[height=0.11\textheight]{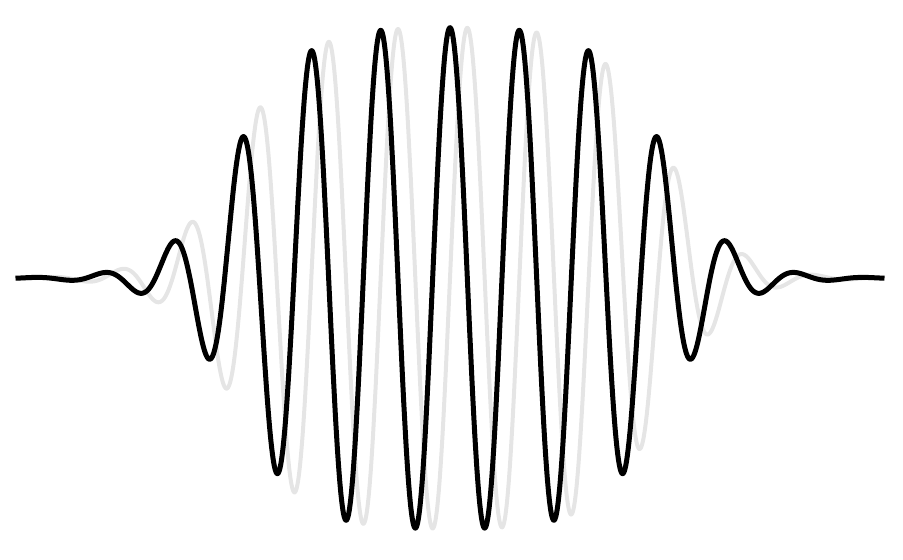}
\includegraphics[height=0.11\textheight]{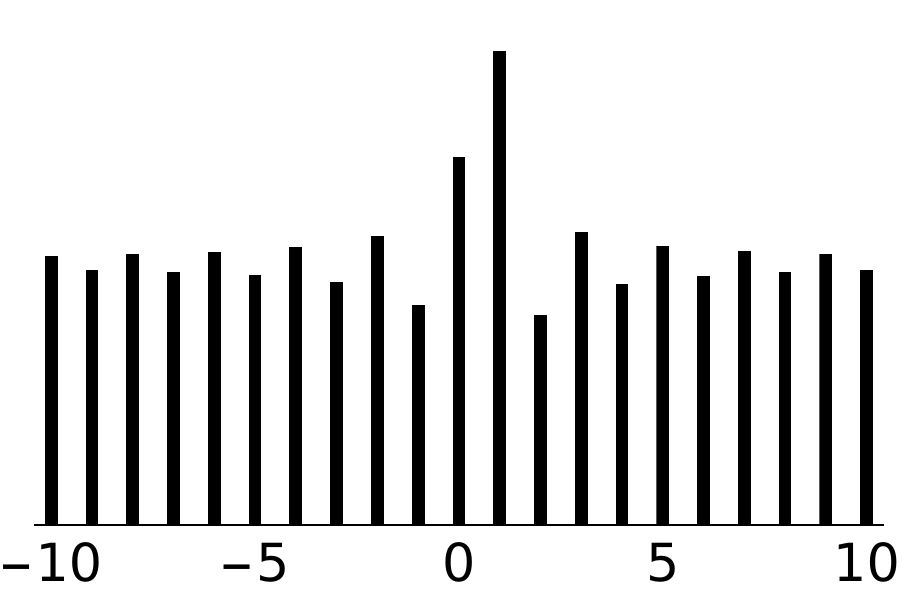}\\
\includegraphics[height=0.11\textheight]{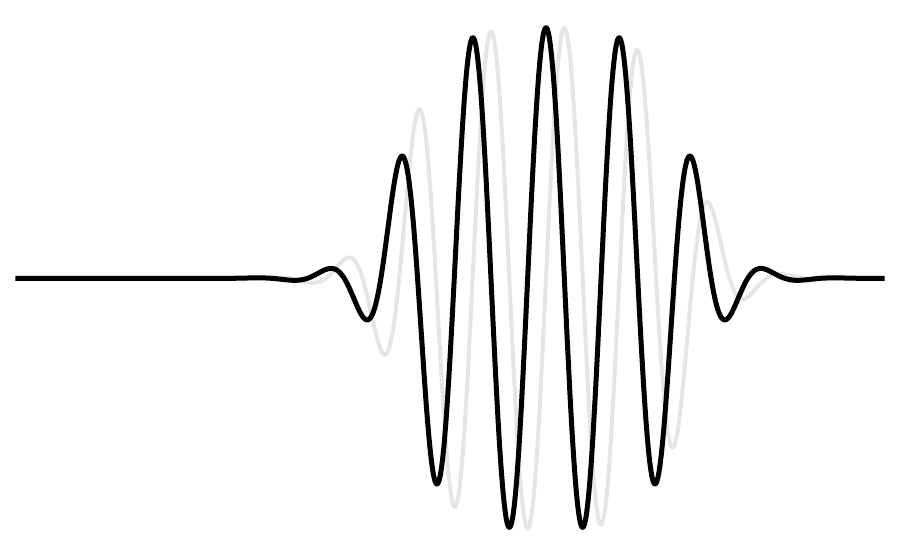}
\includegraphics[height=0.11\textheight]{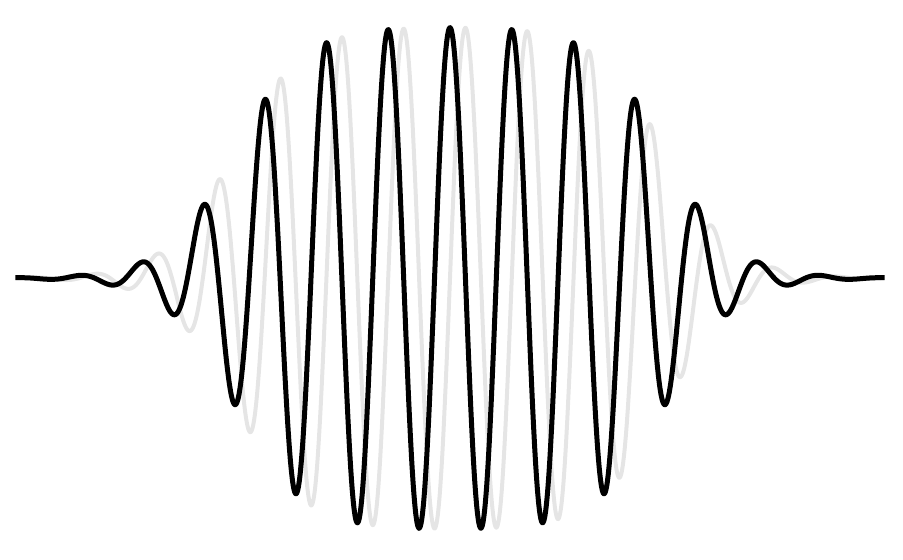}
\includegraphics[height=0.11\textheight]{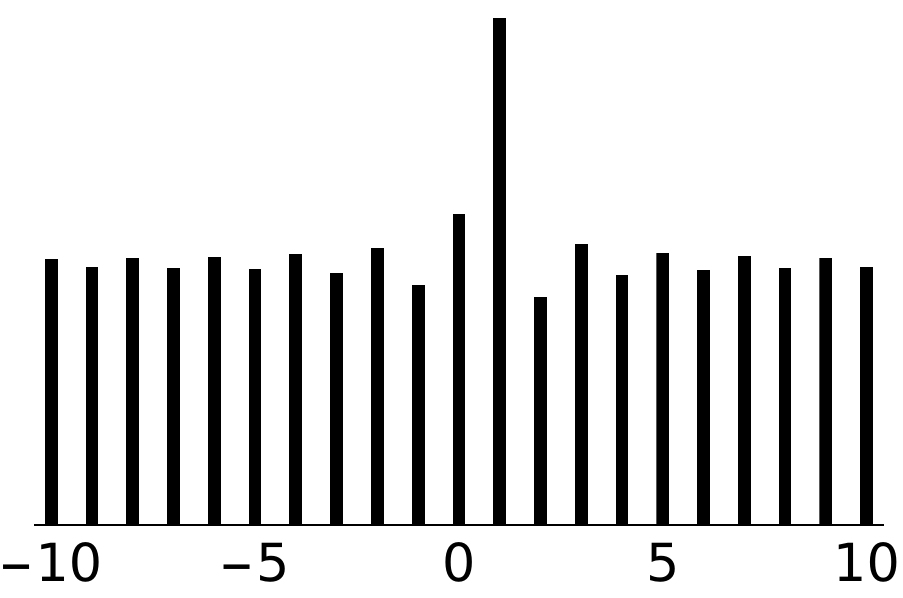}\\
\includegraphics[height=0.11\textheight]{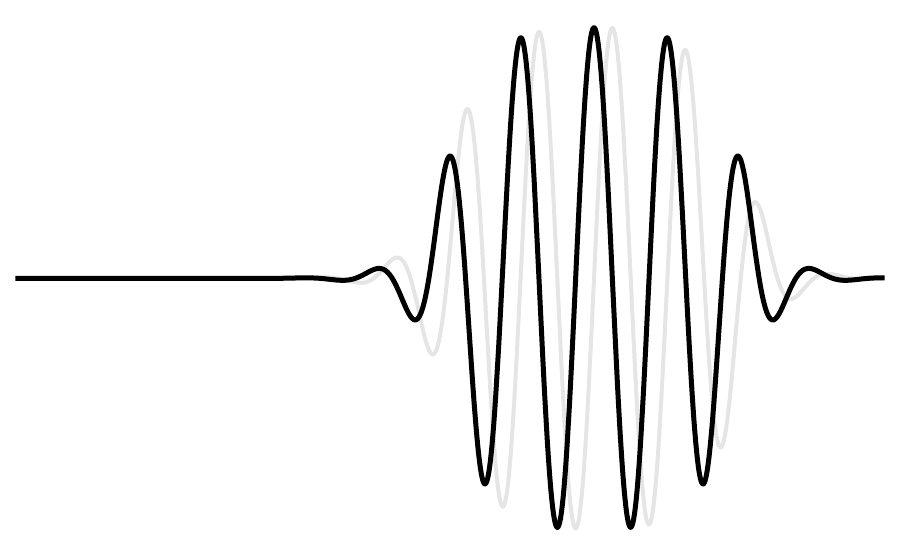}
\includegraphics[height=0.11\textheight]{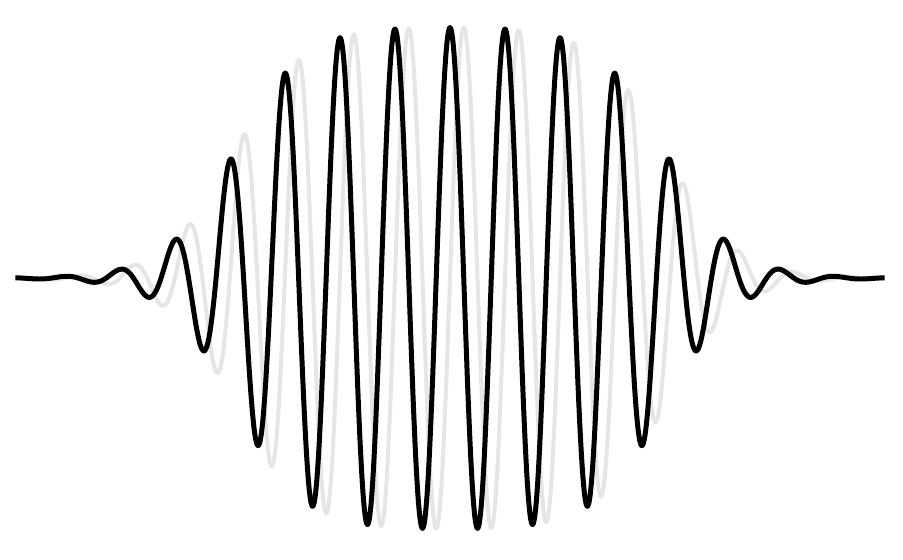}
\includegraphics[height=0.11\textheight]{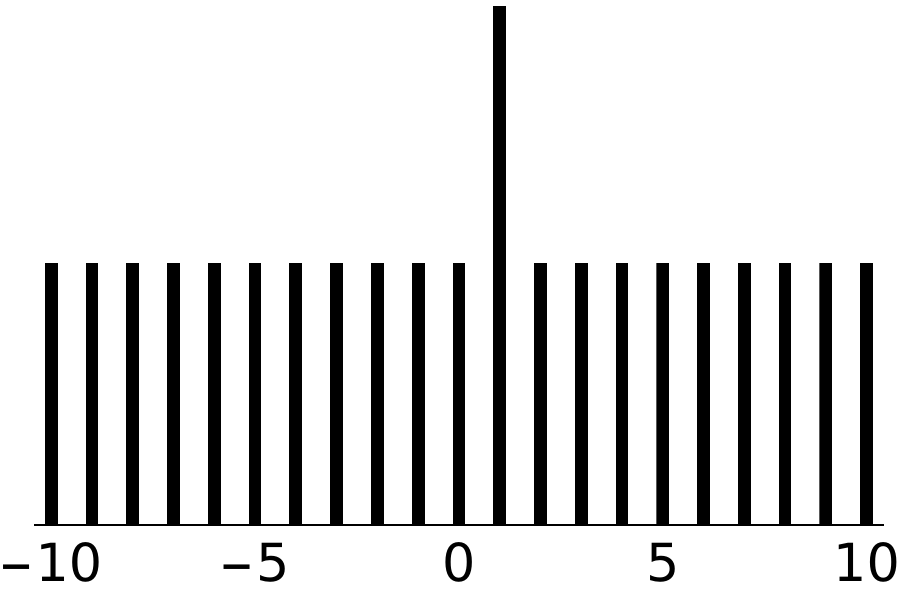}\\
\caption{Three examples of pre-spacetime translation operations (PSTOs) are shown. The first column shows a wave packet being smoothly translated with respect to the old theory's spacetime manifold. The second column shows a wave packet being smoothly increased in frequency. The third column shows one way of smoothly implementing a discrete translation operation on a lattice theory. In all three cases, when viewed as a Lie group action on the theory's states, these smooth transformations are structurally indistinguishable from spacetime translations. Hence, one can use the ISE Methodology to build a new spacetime setting for the theory in question such that these PSTOs become honest-to-goodness spacetime translations in the new theory. }\label{FigPSTOs}
\end{figure}

As a non-trivial example of PSTOs, consider smoothly increasing the frequency of every wave state described by the theory in question. Returning to the above discussed, complex scalar field theory, we might smoothly shift the theory's states as $\varphi_\text{old}(t,x)\mapsto \exp(-\ii\,\Delta k\,x)\,\varphi_\text{old}(t-\Delta t,x)$ for any $\Delta t,\Delta k\in\mathbb{R}$. The action of this Lie group on $\varphi_\text{old}(t,x)$ is shown in the second column of Fig.~\ref{FigPSTOs}. Why do these pitch-shifting transformations qualify as PSTOs? While they are not translations in the old theory's spacetime, they are nonetheless translation operations somewhere, namely in the old theory's Fourier space. Hence, when viewed as a Lie group action on the theory's states, these Fourier shift operations are structurally identical to spacetime translations. Therefore, they qualify as PSTOs.

But what happens when we externalize these Fourier shifts PSTOs? Externalization builds a new spacetime setting for the theory such that these PSTOs become honest-to-goodness spacetime translations in the new theory. Hence, by construction what were translations in Fourier space in the old theory will become translations in spacetime in the new theory. The new theory's spacetime is effectively a copy of the old theory's Fourier space. In this example the ISE Methodology effectively implements a Fourier transform (see Sec.~\ref{SecDemoExt}).

After Secs.~\ref{SecSurInt} and~\ref{SecDemoExt} demonstrate the ISE Methodology in this relatively familiar context Sec.~\ref{SecDemoISE} will use the exact same topological redescription techniques to do something much more interesting. Namely, I will use the ISE Methodology to redescribe a lattice theory (i.e., a theory set on a discrete spacetime, $\mathcal{M}_\text{old}\cong\mathbb{R}\times\mathbb{Z}$) as existing on a continuous spacetime manifold, $\mathcal{M}_\text{new}\cong\mathbb{R}\times\mathbb{R}$. This example application is inspired by the work of \cite{Kempf_2010} entitled ``Spacetime could be simultaneously continuous and discrete, in the same way that information can be'' although I ultimate approach it from a substantially different direction (see Sec.~\ref{SecDemoISE}). On my account this discrete-to-continuous spacetime redescription is facilitated by the fact that the lattice theory in question admits some continuous pre-spacetime translation operations. These continuous PSTOs (shown in the third column of Fig.~\ref{FigPSTOs}) act on the theory's discrete states so as to smoothly interpolate between the theory's discrete spacetime translation. When viewed as a Lie group action on the theory's states, these transformations are structurally identical to spacetime translations. (Namely, they are structurally identical to spacetime translations in exactly the same sense as the above-discussed Fourier shifts were). Hence, they too qualify as PSTOs. 

But what prompts us to take these continuous PSTOs seriously as spacetime translations? As Sec.~\ref{SecDemoISE} will discuss the transformations shown in Fig.~\ref{FigPSTOs} are not only PSTOs but also dynamical symmetries (i.e., they map solutions to solutions). We have thus discovered that these theories have a hidden dynamical symmetry! The ISE Methodology allows us to redescribe these lattice theories in such a way that their hidden dynamical symmetries are re-expressed as spacetime symmetries in the new theory. The process for constructing this new continuous spacetime is identical to the Fourier example discussed above.

\section{Surviving Internalization: Dynamics, Diffeomorphisms, and the Spacetime Manifold}\label{SecSurInt}
The first step of the ISE Methodology is internalization. Let us see it in action. Consider the following non-linear theory set on a curved spacetime:
\begin{quote}
{\hypertarget{QKGSphere}{\bf A Quartic Klein Gordon Theory (2+1D QKG) - }}  Consider a spacetime theory about a scalar field, $\varphi_\text{old}:\mathcal{M}_\text{old}\to\mathcal{V}_\text{old}$, with spacetime, $\mathcal{M}_\text{old}\cong\mathbb{R}\times S^2$, and value space, $\mathcal{V}_\text{old}\cong\mathbb{R}$. For this theory, there exists an injective map, $C:\mathcal{M}_\text{old}\to \mathbb{R}^4$, which assigns to each point, $p\in \mathcal{M}_\text{old}$, coordinates, $C(p)=(t,x,y,z)\in\mathbb{R}^4$ with \mbox{$x^2+y^2+z^2=1$}. This theory's states are subject to the following kinematic constraint: In this fixed coordinate system the field, $\varphi_\text{old}$, must be smooth. In this coordinate system, the metaphysically relevant way to judge the relative size and similarity of this theory's states is with an $L^2$ inner product. In this coordinate system, the theory's states obey the following dynamics:
\begin{align}\label{QKGSphere}
(\partial_t^2-L_x^2-L_y^2-L_z^2+M^2)\,\varphi_\text{old}
+\lambda \, \varphi_\text{old}^3 = 0,
\end{align} 
where $L_x$, $L_y$, and $L_z$ are the generators of rotations about the $x$, $y$, and $z$-axis respectively. $\lambda\in\mathbb{R}$ is some fixed self-coupling constant and $M\geq0$ is some fixed mass parameter.
\end{quote}
Let us begin by identifying the modal structure of this theory. At its broadest level, this theory is about the set, $S_\text{old}^\text{all}$, of all $\mathbb{R}$-valued functions definable on $\mathcal{M}_\text{old}\cong\mathbb{R}\times S^2$ (even those which are non-smooth and/or discontinuous). Enforcing the theory's kinematic constraints, we can identify within $S_\text{old}^\text{all}$ a set of kinematically allowed states, $S_\text{old}^\text{kin}\subset S_\text{old}^\text{all}$. This is the set of all smooth $\mathbb{R}$-valued functions on $\mathcal{M}_\text{old}$. Within this set, we can then enforce the theory's dynamics to find its dynamically allowed states, $S_\text{old}^\text{dyn}\subset S_\text{old}^\text{kin}\subset S_\text{old}^\text{all}$. 

A crucial first step in the internalization process is to identify which structural features of one's theory are dynamically and metaphysically relevant. In particular, the theory's kinematically allowed states, $S_\text{old}^\text{kin}$, must come equipped with all of the resources needed to state the theory's dynamics and to make sense of its metaphysics. For instance, given its dynamics, the Quartic Klein-Gordon theory requires us to define three point-wise operations on $\mathcal{M}_\text{old}$. Namely, it requires an addition operation, $+_\text{old}:S_\text{old}^\text{kin}\times S_\text{old}^\text{kin}\to S_\text{old}^\text{kin}$, and a scalar multiplication operation, $\cdot_\text{old}:\mathbb{R}\times S_\text{old}^\text{kin}\to S_\text{old}^\text{kin}$, and finally a product operation, $\times_\text{old}:S_\text{old}^\text{kin}\times S_\text{old}^\text{kin}\to S_\text{old}^\text{kin}$. Using these we can state the theory's dynamical equations and define its dynamically allowed states, $S_\text{old}^\text{dyn}\subset S_\text{old}^\text{kin}$. Past its dynamically-relevant structure, this theory also comes equipped with a metaphysically relevant inner product, $\langle\varphi,\phi\rangle_\text{old}$, which we can use to judge the relative size and similarity of any two kinematically allowed states, $\varphi,\phi\in S_\text{old}^\text{kin}$. 

Before we can begin internalizing this theory, two more pieces of mathematical structure need to be pointed out. Recall that the goal of the internalization process is to divorce the dynamical behavior of matter from the theory's assumed topological background structure (i.e., its spacetime manifold). But where do the dynamical fields, $\varphi_\text{old}:\mathcal{M}_\text{old}\to\mathcal{V}_\text{old}$, make contact with the spacetime manifold? Note that each of the modal spaces discussed above, $S_\text{old}^\text{dyn}\subset S_\text{old}^\text{kin}\subset S_\text{old}^\text{all}$, is populated by \textit{functions on the spacetime manifold}, $\mathcal{M}_\text{old}$. Namely, our theory comes equipped with an operation, $\text{eval}:S_\text{old}^\text{kin}\times\mathcal{M}_\text{old}\to \mathcal{V}_\text{old}$, which allows us to evaluate the field states at spacetime points, $\text{eval}(\varphi_\text{old},p)=\varphi_\text{old}(p)$. It is this connection which the internalization process aims to sever.


One final piece of structure must be noted before we internalize this theory. Note that the Quartic Klein-Gordon theory, admits a map $*_\text{lift}:d\mapsto d^*$, which can lift the theory's diffeomorphisms, $d\in\text{Diff}(\mathcal{M}_\text{old})$, to act on $S_\text{old}^\text{all}$ as $d^*:S_\text{old}^\text{all}\to S_\text{old}^\text{all}$. Concretely, we have $d^*\varphi_\text{old}\coloneqq\varphi_\text{old}\circ d^{-1}$. We can restrict these lifted diffeomorphisms to act only on $S_\text{old}^\text{kin}$ as $d^*\vert_\text{kin}:S_\text{old}^\text{kin}\to S_\text{old}^\text{kin}$. Next note that there is an intuitive sense in which these lifted diffeomorphisms, $d^*\vert_\text{kin}$, act smoothly on $S_\text{old}^\text{kin}$. But in order to understand the action of $d^*\vert_\text{kin}$ on $\mathcal{S}_\text{old}^\text{kin}$ as being smooth, we must be able to equip this theory's space of kinematically allowed states, $S_\text{old}^\text{kin}$, with some smooth topological structure, $\mathcal{T}_\text{old}$. Hence, in general, we can imagine smoothly varying the theory's states via some Lie group actions (e.g., any of those shown in Fig.~\ref{FigPSTOs}) with $d^*\vert_\text{kin}$ being one example.

We can collect together all of the structures associated with the old theory's space of kinematically allowed states as follows:
\begin{align}
\mathcal{A}_\text{old}^\text{kin}=(S_\text{old}^\text{kin},\mathcal{T}_\text{old},+_\text{old},\,\cdot_\text{old}\,,\times_\text{old},\,\langle,\rangle_\text{old}\,,\text{eval},*_\text{lift},\mathcal{M}_\text{old}).
\end{align}
This theory is now ready to be internalized. The way in which the ISE Methodology divorces the theory's states from the old spacetime setting is similar in spirit to the algebraic approach to spacetime. Let us first discuss what these approaches have in common before seeing how the ISE Methodology goes further. On either approach, we begin by noting that a large portion of the theory's structure can be preserved while breaking the field-to-spacetime connection. In order to break this connection, all we need to do is forget about $\mathcal{M}_\text{old}$ (and relatedly the $\text{eval}$ and $*_\text{lift}$ operations). We can do this via some forgetful operation, $\mathcal{F}_\text{alg}$, which remembers only algebraic structure as follows:
\begin{align}
\mathcal{F}_\text{alg}:\mathcal{A}_\text{old}^\text{kin}\to \mathcal{A}_\text{neutral}^\text{kin}\coloneqq (S_\text{neutral}^\text{kin},\mathcal{T}_\text{neutral},+_\text{neutral},\,\cdot_\text{neutral}\,,\times_\text{neutral},\,\langle,\rangle_\text{neutral}\,).
\end{align}
Three things have changed here. Firstly, the old theory's spacetime manifold, $\mathcal{M}_\text{old}$, has been dropped, together with the $\text{eval}$ and $*_\text{lift}$ operations it supported. Secondly, the set, $S_\text{old}^\text{kin}$, has been replaced with an equal-sized set, $S_\text{neutral}^\text{kin}$, whose elements have no connection with $\mathcal{M}_\text{old}$. Namely, whereas $\varphi_\text{old}\in S_\text{old}^\text{kin}$ is a functions on $\mathcal{M}_\text{old}$, the corresponding element $\bm{\varphi}\coloneqq\mathcal{F}_\text{alg}(\varphi_\text{old})\in S_\text{neutral}^\text{kin}$ is not. Thirdly, the algebraic operations appearing in the above expression (e.g., $+$, $\,\cdot\,$, $\times$, and $\langle,\rangle$) are now defined on $S_\text{neutral}^\text{kin}$ instead of $S_\text{old}^\text{kin}$ such that we have an isomorphism, $\mathcal{A}_\text{neutral}^\text{kin}\cong\mathcal{A}_\text{old}^\text{kin}$, as algebras. Concretely, we have, 
\begin{align}\label{InducedPlusTimes}
\bm{\varphi} \ +_\text{neutral}\ \bm{\phi}
&\coloneqq \mathcal{F}_\text{alg}(\mathcal{F}_\text{alg}^{-1}(\bm{\varphi}) +_\text{old} \mathcal{F}_\text{alg}^{-1}(\bm{\phi})),\\
\nonumber
\alpha\,\cdot_\text{neutral}\,\bm{\varphi} 
&\coloneqq \mathcal{F}_\text{alg}(\alpha\,\cdot_\text{old}\,\mathcal{F}_\text{alg}^{-1}(\bm{\varphi})),\\
\nonumber
\bm{\varphi} \ \times_\text{neutral}\ \bm{\phi}
&\coloneqq\mathcal{F}_\text{alg}(\mathcal{F}_\text{alg}^{-1}(\bm{\varphi}) \times_\text{old} \mathcal{F}_\text{alg}^{-1}(\bm{\phi})),\\
\nonumber
\langle\bm{\varphi},\bm{\phi}\rangle_\text{neutral}
&\coloneqq\langle\mathcal{F}_\text{alg}^{-1}(\bm{\varphi}), \mathcal{F}_\text{alg}^{-1}(\bm{\phi})\rangle_\text{old},
\end{align}
where $\bm{\varphi},\bm{\phi}\in \mathcal{A}_\text{neutral}^\text{kin}$ are generic spacetime-neutral states and $\alpha\in\mathbb{R}$. Similarly, the set of spacetime neutral states $\bm{\varphi}\in S_\text{neutral}^\text{kin}$ has a smooth structure, $\mathcal{T}_\text{neutral}$, which isomorphic to the old theory's, $\mathcal{T}_\text{old}$. Namely, it has the smooth structure which is naturally induced by the bijection, $\mathcal{F}_\text{alg}:S_\text{old}^\text{kin}\leftrightarrow S_\text{neutral}^\text{kin}$. In total, $\mathcal{F}_\text{alg}$ forgets everything about the Quartic Klein-Gordon theory except for its algebraic structure and its smooth topological structure of its kinematically allowed states. 

It is important to note that exactly how forgetful the $\mathcal{F}$ operation ought to be will vary from theory to theory. Ultimately, the correct level of forgetfulness for $\mathcal{F}$ will be set by the dynamical and metaphysical details of the theory in question. Two more examples of internalization will be discussed in this paper. \cite{ISEEquiv} will discuss the internalization process in complete generality as well as giving two more concrete examples. The ISE Methodology allows for a wide variability of the dynamically and metaphysically relevant structures which are at play. There are, however, some core structural features which are \textit{mechanically necessary} for the ISE Methodology to be applied to some theory. 

If one comes from a spacetime algebraicist perspective, one might expect that some degree of algebraic structure (e.g., $+$, $\,\cdot\,$, $\times$, and $\langle,\rangle$) will be required later on in constructing a new spacetime framing of this theory (e.g., if one seeks to identify spacetime points with certain maximal ideals of some ring). It is at this point, however, that the ISE Methodology diverges substantially from spacetime algebraicism. Namely, unlike the algebraic approach to spacetime, the ISE Methodology does not make use of any algebraic structure in recovering the theory's old spacetime settings (or in generating new spacetime settings). Hence, in principle, the internalization process can be extremely forgetful; Concretely, we might internalize via a forgetful operation,
\begin{align}
\mathcal{F}_\text{smooth}:\mathcal{A}_\text{old}^\text{kin}\to \mathcal{S}_\text{neutral}^\text{kin}
=(S_\text{neutral}^\text{kin},\mathcal{T}_\text{neutral}),
\end{align}
which remembers only the smooth structure of the theory's kinematically allowed state space. As noted in Sec.~\ref{SecPreview}, the new spacetime settings offered to us by the ISE Methodology are to be built solely from our ability to smoothly vary the theory's states (recall Fig.~\ref{FigPSTOs}).

To be clear, there may be dynamical or metaphysical reasons for preserving other aspects of the theory in question through the internalization process (e.g., its algebraic structure if it has any). These structures may play a substantial role in stating the theory's dynamics, in grounding a satisfying metaphysical story, and even in choosing the best spacetime setting for the theory. Nonetheless, any such structures are \textit{mechanically irrelevant} throughout the ISE Methodology. They merely come along for the ride so to speak. It is only the smooth structure, $\mathcal{T}$, of the theory's state space which is invoked in the actual construction of new candidate spacetime settings (see Sec.~\ref{SecDemoExt} for details).

Let us return to internalizing the Quartic Klein-Gordon theory, which incidentally does require us to preserve its algebraic structure. What survives internalization? The old theory's kinematically allowed states, $\varphi_\text{old}\in \mathcal{A}_\text{old}^\text{kin}$, survive internalization as follows,
\begin{align}
\varphi_\text{old}\surint 
\bm{\varphi}\coloneqq\mathcal{F}_\text{alg}(\varphi_\text{old}). 
\end{align}
Moreover, given that $\mathcal{A}_\text{old}^\text{kin}$'s algebraic structure has faithfully survived internalization, we ought to have the resources within $\mathcal{A}_\text{neutral}^\text{kin}$ to state the theory's dynamics. Let us see how this happens. To begin, recall the theory's dynamical equation given by Eq.~\eqref{QKGSphere},
\begin{align}
(\partial_t^2-L_x^2-L_y^2-L_z^2+M^2)\,\varphi_\text{old}
+\lambda \, \varphi_\text{old}^3 = 0,
\end{align}
Each part of this dynamical equation survives internalization as follows:
\begin{align}
\partial_t\vert_\text{kin}&\surint& D_0&\coloneqq\mathcal{F}_\text{alg}\circ \partial_t\circ\mathcal{F}_\text{alg}^{-1}\\
\nonumber
L_{x,y,z}\vert_\text{kin}&\surint& D_{1,2,3}&\coloneqq\mathcal{F}_\text{alg}\circ L_{x,y,z}\circ\mathcal{F}_\text{alg}^{-1}
\end{align}
Hence, we have the dynamical equation itself surviving internalization as follows:
\begin{align}
&(\partial_t^2-L_x^2-L_y^2-L_z^2+M^2)\,\varphi_\text{old}
+\lambda \, \varphi_\text{old}^3 = 0\\
\nonumber
&\surint
(D_0^2-D_1^2-D_2^2-D_3^2+M^2)\,\bm{\varphi}
+\lambda \cdot\bm{\varphi}^3 = 0
\end{align}
where the $\,\cdot\,$, $+$, and $\times$ operations in the spacetime neutral theory are those induced by the forgetful operation, $\mathcal{F}_\text{alg}$, see Eq.~\eqref{InducedPlusTimes}. Ultimately, this means that we can pick out the theory's dynamically allowed states in a spacetime-neutral way. Hence, this space too survives internalization,
\begin{align}
\mathcal{S}_\text{old}^\text{dyn}\subset \mathcal{A}_\text{old}^\text{kin}
&\surint
\mathcal{S}_\text{neutral}^\text{dyn}\subset \mathcal{A}_\text{neutral}^\text{kin}.
\end{align}

We have so far seen how a theory's states and dynamics can survive being divorced from the old theory's topological background structure. Importantly, however, internalization does not simply delete all of the theory's topological information. Surprisingly, a great deal of topological information about the old spacetime setting can survive internalization. The key to seeing how any topological information survives internalization the $*_\text{lift}:d\mapsto d^*$ map discussed above. Recall that this map lifts diffeomorphisms, $d:\mathcal{M}_\text{old}\to \mathcal{M}_\text{old}$, to act on the theory's states as $d^*:S_\text{old}^\text{all}\to S_\text{old}^\text{all}$. Once we restrict the action of $d^*$ to kinematically allowed states  (e.g., $d^*\vert_\text{kin}:S_\text{old}^\text{kin}\to S_\text{old}^\text{kin}$), it becomes possible to see it surviving internalization. That is, we can see them as maps on $S_\text{neutral}^\text{kin}$, namely $G(d)\coloneqq \mathcal{F}_\text{alg} \circ d^*\vert_\text{kin} \circ \mathcal{F}_\text{alg}^{-1}$.

For the Quartic Klein-Gordon theory, a notable family of diffeomorphism on $\mathcal{M}_\text{old}\cong\mathbb{R}\times S^2$ survive internalization in this way. Let $d_t(\epsilon)\in\text{Diff}(\mathcal{M}_\text{old})$ be the diffeomorphism on $\mathcal{M}_\text{old}$ which act as $d_t(\epsilon):(t,x,y,z)\mapsto (t+\epsilon,x,y,z)$ in the $C$-coordinate system. Similarly, let $d_x(\theta), d_y(\theta), d_z(\theta)\in\text{Diff}(\mathcal{M}_\text{old})$, be the diffeomorphisms which rigidly rotate $\mathcal{M}_\text{old}$ around the $x$, $y$ and $z$-axis  in the $C$-coordinate system. Let $H_\text{trans}\cong (\mathbb{R},+)\times\text{SO}(3)$ be the Lie group formed by these diffeomorphisms. Each of these diffeomorphisms, $d\in H_\text{trans}$ survives internalization as follows:
\begin{align}
d_t^*(\epsilon)\vert_\text{kin} = \text{exp}(-\epsilon\,\partial_t) \surint G(d_t(\epsilon)) = \text{exp}(-\epsilon \, D_0)\\
\nonumber
d_x^*(\theta)\vert_\text{kin} = \text{exp}(-\theta L_x) \surint G(d_x(\theta)) = \text{exp}(-\theta D_1)\\
\nonumber
d_y^*(\theta)\vert_\text{kin} = \text{exp}(-\theta L_y) \surint G(d_y(\theta)) = \text{exp}(-\theta D_2)\\
\nonumber
d_z^*(\theta)\vert_\text{kin} = \text{exp}(-\theta L_z) \surint G(d_z(\theta)) = \text{exp}(-\theta D_3).
\end{align}
Indeed, these diffeomorphisms survive internalization not only individually, but also as a Lie group (i.e., the $G$ map is a Lie group isomorphism). In total, therefore, we have that the Lie group $H_\text{trans}\subset\text{Diff}(\mathcal{M}_\text{old})$ faithfully survives internalization as follows:
\begin{align}\label{HtransGtrans}
H_\text{trans}\surint G_\text{trans}\coloneqq G(H_\text{trans})
\end{align}
with $G_\text{trans}\cong H_\text{trans}\cong (\mathbb{R},+)\times\text{SO}(3)$. Thus, at least some topological information regarding the old spacetime setting, $\mathcal{M}_\text{old}$, has survived internalization. 

Perhaps surprisingly, not just partial but complete topological information about $\mathcal{M}_\text{old}$ has faithfully survived internalization. Seeing this, however, will require that we first understand a bit of mathematics regarding the quotients of Lie groups, namely, smooth homogeneous manifolds. A summary of this topic along with proofs of some key results can be found in Appendix~\ref{AppHomoMan}. In order to see $\mathcal{M}_\text{old}$ as surviving internalization we will first reconstruct it as a quotient of its diffeomorphisms. To do this, we need a pick out a finite-dimensional Lie group, $H\subset\text{Diff}(\mathcal{M}_\text{old})$, which acts transitively on $\mathcal{M}_\text{old}$. The Lie group $H_\text{trans}\cong (\mathbb{R},+)\times\text{SO}(3)$ discussed above is just such a Lie group (hence the subscript ``trans'').  By using the \hyperlink{HMChaThm}{Homogeneous Manifold Characterization Theorem} (proved in Appendix~\ref{AppHomoMan}) we can reconstruct $\mathcal{M}_\text{old}$ up to diffeomorphism as,
\begin{align}\label{MoldHH}
\mathcal{M}_\text{old}\cong \mathcal{M}_\text{H}\coloneqq \frac{H_\text{trans}}{H_\text{fix}}\cong \frac{(\mathbb{R},+)\times\text{SO}(3)}{\{\openone\}\times\text{SO}(2)},
\end{align}
where $H_\text{fix}\cong \{\openone\}\times\text{SO}(2)$ is one of $H_\text{trans}$'s stabilizer subgroups. A concrete demonstration of the non-trivial part of this quotient, $\text{SO}(3)/\text{SO}(2)\cong S^2$, can be found in Appendix~\ref{AppHomoMan}.

Having reconstructed the old theory's spacetime manifold in this way, we can now see it as faithfully surviving internalization. As noted above, the Lie group $H_\text{trans}$ (and, consequently, $H_\text{fix}\subset H_\text{trans}$) faithfully survives internalization:
\begin{align}
H_\text{trans}
&\surint G_\text{trans}\coloneqq G(H_\text{trans})\cong H_\text{trans},\\
\nonumber
H_\text{fix}
&\surint G_\text{fix}\coloneqq G(H_\text{fix})\cong H_\text{fix},
\end{align}
becoming some Lie groups, $G_\text{fix}\subset G_\text{trans}$, acting on $\mathcal{A}_\text{neutral}^\text{kin}$. By taking the quotient of these surviving Lie groups we can faithfully reconstruct $\mathcal{M}_\text{H}$ (and, consequently, $\mathcal{M}_\text{old}\cong\mathcal{M}_\text{H}$) post-internalization as follows:
\begin{align}\label{ManSur1}
\mathcal{M}_\text{old}\cong\mathcal{M}_\text{H}\coloneqq \frac{H_\text{trans}}{H_\text{fix}}
\surint
\mathcal{M}_\text{G}\coloneqq \frac{G_\text{trans}}{G_\text{fix}}\cong\frac{H_\text{trans}}{H_\text{fix}}\cong \mathcal{M}_\text{old}.
\end{align}
Thus, the Quartic Klein-Gordon theory's spacetime manifold has faithfully survived internalization. 

This is a surprising conclusion. Recall that the stated purpose of internalization was to conceptually divorce the theory's states, $\varphi_\text{old}$, from any assumed topological background structure. Despite this divorce, it seems that complete topological information about $\mathcal{M}_\text{old}\cong\mathbb{R}\times S^2$ has survived internalization. As~\cite{ISEEquiv} proves, for any theory with a sufficient level of spacetime-kinematic compatibility, there is enough topological information contained in the surviving Lie groups $G_\text{trans}$ and $G_\text{fix}$ to perfectly reconstruct the theory's original spacetime setting. Namely,  by externalizing these Lie groups, one can reconstruct the old theory's spacetime setting, up to diffeomorphism. 

Before discussing the externalization process, however, allow me to first demonstrate how the internalization process applies to one more example theory. Consider the following theory about a modified quantum harmonic oscillator (QHO):
\begin{quote}
\hypertarget{NLQHO}{{\bf Non-Local QHO -}} Consider a spacetime theory about a scalar field, $\varphi_\text{old}:\mathcal{M}_\text{old}\to\mathcal{V}_\text{old}$, with spacetime, \mbox{$\mathcal{M}_\text{old}\cong\mathbb{R}^2$}, and value space, $\mathcal{V}_\text{old}\cong\mathbb{C}$. This theory's states are subject to the following kinematic constraint: In some fixed global coordinate system, $(t,x)$, each state must be within the usual rigged Hilbert space of planewaves and Dirac deltas. In this coordinate system, the metaphysically relevant way to judge the relative size and similarity of this theory's states is with an $L^2$ inner product. In this coordinate system, the theory's states obey the following dynamics:
\begin{align}
\nonumber
\ii\partial_t\varphi_\text{old}(t,x)
&=(-\partial_x^2+x^2)\varphi_\text{old}(t,x)+ \frac{\lambda}{2} \varphi_\text{old}(t,x-a)+ \frac{\lambda}{2} \varphi_\text{old}(t,x+a),
\end{align}
for some fixed $a>0$ and $\lambda\in\mathbb{R}$.
\end{quote}
Note that this is the standard QHO plus a non-local self-coupling.

As with the Quartic Klein-Gordon theory, we ought to begin by identifying this theory's dynamically and metaphysically relevant features. Since this theory's dynamical equations are linear we need the space of kinematically allowed states, $V_\text{old}^\text{kin}$, to have at least a vector space structure. The 
space of dynamically allowed states is then a vector subspace thereof, $V_\text{old}^\text{dyn}\subset V_\text{old}^\text{kin}$. Past its dynamically-relevant linear structure, this theory also has a metaphysically-relevant inner product, $\langle \varphi,\phi\rangle_\text{old}$, defined for $\varphi,\phi\in V_\text{old}^\text{kin}$. As in the previous case, we can internalize this theory by applying a forgetful operation, $\mathcal{F}_\text{vec}$, at the level of kinematics which  only remembers $V_\text{old}^\text{kin}$'s vector space structure as follows:
\begin{align}\label{EqQHOSurvival}
\mathcal{F}_\text{vec}:V_\text{old}^\text{kin}\to V_\text{neutral}^\text{kin}\coloneqq (S_\text{neutral}^\text{kin},\mathcal{T}_\text{neutral},+_\text{neutral},\,\cdot_\text{neutral}\,,\,\langle,\rangle_\text{neutral}).
\end{align}

To finish internalizing this theory, let us see how its spacetime manifold faithfully survives internalization. If $\mathcal{M}_\text{old}$ is to survive internalization we must first reconstruct it as a quotient of its diffeomorphisms. To do this, we need a pick out a finite-dimensional Lie group, $H_\text{trans}\subset\text{Diff}(\mathcal{M}_\text{old})$, which acts transtively $\mathcal{M}_\text{old}$. According to the \hyperlink{HMChaThm}{Homogeneous Manifold Characterization Theorem} it does not matter which Lie group, $H_\text{trans}$, we pick (see Appendix~\ref{AppHomoMan}). The simplest Lie group which satisfies these criteria is the group, $H_\text{trans}\cong(\mathbb{R}^2,+)$, which implements rigid translations in the old theory's $(t,x)$ coordinate system. Note that the stabilizer subgroups for this choice of $H_\text{trans}$ are all identical, $H_\text{fix}=\{\openone\}$.

Whichever Lie groups we use to reconstruct the old theory's spacetime manifold, $\mathcal{M}_\text{old}\cong H_\text{trans}/H_\text{fix}$, we can subsequently see it surviving internalization as follows. The Lie groups in the numerator and denominator of this quotient each faithfully survive internalization becoming a pair of Lie groups, $G_\text{trans}\cong H_\text{trans}$ and $G_\text{fix}\cong H_\text{fix}$, which act on $V_\text{neutral}^\text{kin}$, the space of kinematically allowed states in the spacetime-neutral theory. Hence, the theory's old spacetime manifold, $\mathcal{M}_\text{old}\cong H_\text{trans}/H_\text{fix}$, also faithfully survives internalization becoming $\mathcal{M}_\text{G}\coloneqq G_\text{trans}/G_\text{fix}\cong \mathcal{M}_\text{old}$.


\section{Demonstrating Externalization in a Familiar Setting: Fourier Redescription}\label{SecDemoExt} 
The previous section discussed how a theory's original spacetime manifold, $\mathcal{M}_\text{old}$, can faithfully survive internalization. Namely, it can survive as a quotient of two Lie groups, $G_\text{trans}^\text{(int)}$ and $G_\text{fix}^\text{(int)}$, acting smoothly on $\mathcal{S}_\text{neutral}^\text{kin}$ (e.g., either $\mathcal{A}_\text{neutral}^\text{kin}$ or $V_\text{neutral}^\text{kin}$). The superscript (int) here indicates that these Lie groups were produced by the internalization process. Of course, knowing how a theory's spacetime manifold survives internalization is very suggestive of how externalization might be able to generate a new spacetime manifold for this theory. 

This section will introduce the externalization process by using it to build a new spacetime setting for the non-local QHO theory discussed above. In particular, the new theory's spacetime will be effectively a copy of the old theory's Fourier space and its states will be related to the old theory's states via a Fourier transform. While it may seem like overkill to use the ISE Methodology to achieve this Fourier redescription, the techniques demonstrated here are applicable much more generally. See, for instance, Sec.~\ref{SecDemoISE} as well as two additional applications of the ISE Methodology in \cite{ISEEquiv}. In fact, it is there proved that the ISE Methodology is a completely general tool for topological redescription; Using it one can access every possible spacetime framing of a given theory, bounded only by a weak spacetime-kinematic compatibility condition.

Without further ado, let us see the externalization process in action. Externalization builds new spacetime settings from certain building blocks, namely, pre-spacetime translation operations (PSTOs). Concretely, these are a pair of Lie group, $G_\text{trans}^\text{(ext)}$ and $G_\text{fix}^\text{(ext)}$, which act smoothly on $\mathcal{S}_\text{neutral}^\text{kin}$. Not just any pair of Lie group actions will qualify as PSTOs, however, they must be, in a certain sense, structurally indistinguishable from spacetime translations.
\begin{quote}
{\hypertarget{PSTO}{\bf Definition: Pre-Spacetime Translation Operations (PSTOs) - }} Let $G_2\subset G_1$ be a pair of Lie groups which act smoothly on some space $\mathcal{S}_\text{N}=(S_\text{N},\mathcal{T}_\text{N},\dots)$. This pair of Lie group actions will be called \textit{pre-spacetime translation operations} when they have both Faithful Construction Compatibility and Injection Compatibility.
\end{quote}
The technical definitions of faithful construction compatibility and injection compatibility will be given throughout this section. Roughly, faithful construction compatibility guarantees that our PSTOs, $G_\text{trans}^\text{(ext)}$, will end up faithfully represented as diffeomorphisms on the new spacetime setting, namely, as $H_\text{trans}^\text{(ext)}\subset\text{Diff}(\mathcal{M}_\text{new})$ with $H_\text{trans}^\text{(ext)}\cong G_\text{trans}^\text{(ext)}$. Injection compatibility will then guarantee that the Lie group action of $G_\text{trans}^\text{(ext)}$ on $\mathcal{S}_\text{neutral}^\text{kin}$ is faithfully represented by the lifted actions these diffeomorphisms, $H_\text{trans}^\text{(ext)}{}^*$ on some yet-to-be-defined states on $\mathcal{M}_\text{new}$. This justifies the following gloss of the externalization process: Externalization builds a new spacetime by taking some hand-picked pre-spacetime translation operations, $G_\text{fix}^\text{(ext)}\subset G_\text{trans}^\text{(ext)}$, and letting them be honest-to-goodness spacetime translations, $H_\text{fix}^\text{(ext)}\subset H_\text{trans}^\text{(ext)}\subset\text{Diff}(\mathcal{M}_\text{new})$.

Let us begin by briefly revisiting the example PSTOs discussed in Sec.~\ref{SecPreview} (and displayed in Fig.~\ref{FigPSTOs}). In Sec.~\ref{SecPreview} I claimed that the old theory's spacetime translation operations will always qualify as PSTOs (see the first column of Fig.~\ref{FigPSTOs}). More technically, this claim is about the corresponding Lie group actions which have faithfully survived internalization; Namely, $G_\text{trans}^\text{(int)}$ and $G_\text{fix}^\text{(int)}$ acting on $\mathcal{S}_\text{neutral}^\text{kin}$. \cite{ISEEquiv} proves that this pair of Lie group actions will qualify as PSTOs whenever the theory in question has a sufficient level of spacetime-kinematic compatibility. Moreover, it is there proved that by externalizing these PSTOs one can easily return to the theory's original spacetime setting.

Importantly, however, we are not forced to go back the way we came. We are free to search for other pre-spacetime translation operations to externalize, i.e., ones with $G_\text{trans}^\text{(ext)}\neq G_\text{trans}^\text{(int)}$ and/or $G_\text{fix}^\text{(ext)}\neq G_\text{fix}^\text{(int)}$. To have a concrete alternative in mind, consider the non-local QHO theory introduced in Sec.~\ref{SecSurInt}. Let the maps,
\begin{align}
f(\Delta t,\Delta k)\,\varphi_\text{old}(t,x)=e^{-\ii\Delta k x}\,\varphi_\text{old}(t-\Delta t,x),
\end{align}
form a Lie group $f(\Delta t,\Delta k)\in H_\text{Fourier}$ which acts on $V_\text{old}^\text{kin}$. Some smooth transformations of this kind are depicted in the second column of Fig.~\ref{FigPSTOs}. Note that these $H_\text{Fourier}$ transformations faithfully survive internalization becoming a Lie group, $G_\text{Fourier}=\mathcal{F}_\text{vec}\circ H_\text{Fourier}\circ \mathcal{F}^{-1}_\text{vec}$, which acts on $V_\text{neutral}^\text{kin}$.

There is an intuitive sense in which $G_\text{Fourier}$'s Lie group action on $V_\text{neutral}^\text{kin}$ should qualify as PSTOs; Namely, when viewed as a Lie group action on the theory's states, these Fourier shift operations are structurally identical to spacetime translations. Indeed, while these aren't translations in the old theory spacetime, they are nonetheless translations somewhere, namely in the old theory's Fourier space. Intuitively, we should expect that externalizing the PSTOs ought to produce a new theory which is, roughly speaking, set on a copy of the old theory's Fourier space. With this example of PSTOs in mind, let us next see how the externalization process works more generally.

Recall from Sec.~\ref{SecSurInt} that a theory's spacetime manifold can faithfully survive internalization as a quotient of two Lie groups, $G_\text{trans}^\text{(int)}$ and $G_\text{fix}^\text{(int)}$, acting on $\mathcal{S}_\text{neutral}^\text{kin}$. Analogously, externalization begins from a pair of Lie groups $G_\text{trans}^\text{(ext)}$ and $G_\text{fix}^\text{(ext)}$, acting on $\mathcal{S}_\text{neutral}^\text{kin}$. Externalization puts forward the quotient manifold of these Lie groups as the theory's new spacetime manifold,
\begin{align}\label{MGdef}
\mathcal{M}_\text{new}\cong\mathcal{M}_\text{G}\coloneqq G_\text{trans}^\text{(ext)}/G_\text{fix}^\text{(ext)},
\end{align}
at least up to diffeomorphism. I say here ``up to diffeomorphism'' because the quotient manifold $\mathcal{M}_\text{G}$ contains excess structure; Its elements are sets of maps on $\mathcal{S}_\text{neutral}^\text{kin}$ as well as manifold points on $\mathcal{M}_\text{G}$. We can remove this excess by applying a forgetful operation $\mathcal{F}_\text{man}$ to $\mathcal{M}_\text{G}$. Namely, this is a diffeomorphism \mbox{$\mathcal{F}_\text{man}:\mathcal{M}_\text{G}\to \mathcal{M}_\text{new}$} where $\mathcal{M}_\text{new}$ is some smooth manifold which is isomorphic to $\mathcal{M}_\text{G}$, i.e., with \mbox{$\mathcal{M}_\text{new}\cong \mathcal{M}_\text{G}$}. In terms of this forgetful operation, the new spacetime manifold is defined as,
\begin{align}\label{EqMFGG}
\mathcal{M}_\text{new}\coloneqq \mathcal{F}_\text{man}(G_\text{trans}^\text{(ext)}/G_\text{fix}^\text{(ext)}).
\end{align}
It is worth noting that the new theory's spacetime, $\mathcal{M}_\text{new}$, ultimately gets its smooth topological structure from the smoothness of $G_\text{trans}^\text{(ext)}$'s action on $\mathcal{S}_\text{neutral}^\text{kin}$. For notational convenience, let us now drop the (ext) superscripts.


In addition to giving $\mathcal{M}_\text{new}$ its smooth structure, the Lie group $G_\text{trans}$ is also naturally represented as diffeomorphisms on $\mathcal{M}_\text{new}$. To see this, note that the left action of $G_\text{trans}$ on itself can be seen as smoothly permuting its $G_\text{fix}$-cosets (that is, smoothly permuting the points $[g_0]\in\mathcal{M}_\text{G}$). Concretely, we have that $g\in G_\text{trans}$ acts smoothly on $\mathcal{M}_\text{G}$ as $\bar{g}([g_0])\coloneqq [g\,g_0]$. These maps $\bar{g}:\mathcal{M}_\text{G}\to \mathcal{M}_\text{G}$ are diffeomorphisms on $\mathcal{M}_\text{G}$ and can be transferred over into diffeomorphisms on $\mathcal{M}_\text{new}$ as,
\begin{align}\label{Hdef}
H(g)\coloneqq\mathcal{F}_\text{man}\circ\bar{g}\circ\mathcal{F}_\text{man}^{-1}.    
\end{align}
In total therefore we have a Lie group homomorphism, \mbox{$H:G_\text{trans}\to \text{Diff}(\mathcal{M}_\text{new})$}, which associates to every element, $g\in G_\text{trans}$, a diffeomorphism, $H(g)\in\text{Diff}(\mathcal{M}_\text{new})$. The image of this map, \mbox{$H_\text{trans}^\text{(ext)}\coloneqq H(G_\text{trans})$}, is a representation of $G_\text{trans}$ as diffeomorphisms on $\mathcal{M}_\text{new}$. 

Recall that our goal is to build a new spacetime setting for the theory in question such that our PSTOs end up as honest-to-goodness spacetime translations on $\mathcal{M}_\text{new}$. To achieve this, we need some assurance that the Lie group, $H_\text{trans}^\text{(ext)}$, will be a faithful representation of our PSTOs, namely that
$H_\text{trans}^\text{(ext)}\cong G_\text{trans}$. As is proved in Appendix~\ref{AppHomoMan}, this will happen if and only if $G_\text{trans}$ and $G_\text{fix}$ have faithful construction compatibility.
\begin{quote}
{\bf Definition: Faithful Construction Compatibility - } Let $G_1$ and $G_2$ be two Lie groups. These two Lie groups have \textit{faithful construction compatibility} if and only if $G_2$ is both a closed subgroup of $G_1$ as well as a core-free subgroup of $G_1$. (N.b., $G_2$ is a core-free subgroup of $G_1$ if and only if $G_2$ contains no non-trivial normal subgroups of $G_1$.)
\end{quote}

Let us now return our attention to externalizing the non-local QHO theory with the following choice of PSTOs, $G_\text{trans}=G_\text{Fourier}$. In order to complete our choice of PSTOs, however, we also need to pick out a Lie subgroup, $G_\text{fix}\subset G_\text{trans}$, which has faithful construction compatibility with $G_\text{trans}$. What are our options?  Note that faithful construction compatibility requires $G_\text{fix}$ to be a closed core-free subgroup of $G_\text{trans}=G_\text{Fourier}$. Next, note that $G_\text{Fourier}\cong(\mathbb{R}^2,+)$ is abelian and hence all of its subgroups are normal. This means that the only core-free subgroup of $G_\text{Fourier}$ is the trivial group. Hence, given our choice of $G_\text{trans}=G_\text{Fourier}$ we are forced to take $G_\text{fix}=\{\openone\}$.

Given these PSTOs, the new theory's spacetime manifold is,
\begin{align}
\mathcal{M}_\text{new}\cong G_\text{trans}/G_\text{fix}\cong\mathbb{R}^2.    
\end{align}
Note that while this new spacetime is diffeomorphic to the old one, $\mathcal{M}_\text{new}\cong \mathcal{M}_\text{old}\cong\mathbb{R}^2$ it is nonetheless distinct, $\mathcal{M}_\text{new}\neq \mathcal{M}_\text{old}$. To see the distinction, recall that Fourier shifts in the old theory, $f(\Delta t,\Delta k)\in H_\text{Fourier}$, have faithfully survived internalization becoming 
$g(\Delta t,\Delta k)\in G_\text{Fourier}=G_\text{trans}$ acting on $V_\text{neutral}^\text{kin}$.
But this is the Lie group which we have just externalized. Thus, post-externalization we will have $g(\Delta t,\Delta k)$ becoming a diffeomorphism, $h(\Delta t,\Delta k)\coloneqq H(g(\Delta t,\Delta k))\in H_\text{trans}^\text{(ext)}\subset\text{Diff}(\mathcal{M}_\text{new})$, on our new two-dimensional manifold. Let $(\tau,q)$ be a coordinate system for $\mathcal{M}_\text{new}\cong\mathbb{R}^2$ such that $h(\Delta t,\Delta k)(\tau,q)=(\tau+\Delta t,q+\Delta k)$. Thus, shifts in Fourier space in the old theory become shifts in spacetime in the new theory. As promised, externalization has turned our choice of $G_\text{trans}=G_\text{Fourier}$ into honest-to-goodness spacetime translations on $\mathcal{M}_\text{new}$.

The connection between $\mathcal{M}_\text{new}$ and the old theory's Fourier space will only become stronger as we proceed to the next step in the externalization process, namely injecting the old theory's states and dynamics into the new spacetime setting. Externalization must somehow map the theory's spacetime-neutral states, $\bm{\varphi}\in V_\text{neutral}^\text{kin}$, and their dynamics onto $\mathcal{M}_\text{new}$. Let us assume that the new theory will be about a scalar field, $\varphi_\text{new}:\mathcal{M}_\text{new}\to\mathcal{V}_\text{new}$ for some value space, $\mathcal{V}_\text{new}$. In order to begin identifying the new theory's states we must first build $S_\text{new}^\text{all}$, the set of all $\mathcal{V}_\text{new}$-valued fields definable on $\mathcal{M}_\text{new}$ (even those which are non-smooth and/or discontinuous). Within this wide-scope we must next pick out the subset of states which are kinematically allowed in the new theory, $S_\text{new}^\text{kin}\subset S_\text{new}^\text{all}$. Recall that we already have a spacetime-neutral characterization of the theory's kinematically allowed states, namely, $S_\text{neutral}^\text{kin}$. All we need to do is to map these states onto the new spacetime setting via some map $J:\bm{\varphi}\mapsto \varphi_\text{new}$. Demanding that distinct states, $\bm{\varphi}\in S_\text{neutral}^\text{kin}$, are mapped onto distinct states, $\varphi_\text{new}\in S_\text{new}^\text{all}$, we are looking for an injective map, \mbox{$J:S_\text{neutral}^\text{kin}\to S_\text{new}^\text{all}$}. Fixing this injector, $J$, will fix the new theory's kinematically allowed states as $S_\text{new}^\text{kin}\coloneqq J(S_\text{neutral}^\text{kin})$. 

Given the above-discussed connection between the new theory's spacetime and the old theory's Fourier space, an intuitive choice of $J$ would be the one which implements a Fourier transform. Namely, we should be able to externalize via an injective map $J$ which (together with $\mathcal{F}_\text{vec}$) defines $\varphi_\text{new}$ as follows:
\begin{align}\label{IntuitiveJFourier}
J\circ\mathcal{F}_\text{vec}:\varphi_\text{old}\mapsto
\varphi_\text{new}(\tau,q)&=\tilde{\varphi}_\text{old}(\tau,q).
\end{align}
That is, the value of  $\varphi_\text{new}$ at coordinate $(\tau,q)$ is the value of $\varphi_\text{old}(t,x)$'s Fourier transform (namely, $\tilde{\varphi}_\text{old}(t,k)$) at $(t,k)=(\tau,q)$. At this point one may rightly wonder, are we somehow forced into this choice of $J$ or are there other ways in which we could externalize this theory onto $\mathcal{M}_\text{new}$? I will return to this question at the end of this section. For now, however, let us proceed with this intuitive choice of $J$ and see where it leads us.

With this choice of injector \mbox{$J:S_\text{neutral}^\text{kin}\to S_\text{new}^\text{all}$}, one can straight-forwardly translate the old theory's dynamics and kinematics onto  its new spacetime setting as follows.
\begin{quote}
\hypertarget{CosQHO}{{\bf Cosine QHO -}} 
Consider a spacetime theory about a scalar field, $\varphi_\text{new}:\mathcal{M}_\text{new}\to\mathcal{V}_\text{new}$, with spacetime, \mbox{$\mathcal{M}_\text{new}\cong\mathbb{R}^2$}, and value space, $\mathcal{V}_\text{new}\cong\mathbb{C}$. The theory's states are subject to the following kinematic constraint: In some fixed global coordinate system, $(\tau,q)$, each state must be within the usual rigged Hilbert space of planewaves and Dirac deltas. In this coordinate system, the metaphysically relevant way to judge the relative size and similarity of this theory's states is with an $L^2$ inner product. In this coordinate system, the theory's states obey the following dynamics:
\begin{align}
\ii\partial_\tau\varphi_\text{new}(\tau,q)
&=(-\partial_q^2+q^2+\lambda\,\text{cos}(a\,q))\,\varphi_\text{new}(\tau,q),
\end{align}
for some $a>0$ and $\lambda\in\mathbb{R}$.
\end{quote}
Note that this is the standard QHO plus a cosine-shaped potential. Note that whereas the old theory's dynamics was non-local, this new theory has local dynamics. In this way, this new spacetime setting is a better fit for this theory's dynamics than the old one was.

To finalize the externalization process, let us see how the old theory's dynamically and metaphysically structures have survived into the new theory. Recall from Sec.~\ref{SecSurInt} (specifically, Eq.~\eqref{EqQHOSurvival}) that the following features of the non-local QHO theory survived internalization: an addition operation (which was point-wise on $\mathcal{M}_\text{old}$), an scalar multiplication operation (which was also point-wise on $\mathcal{M}_\text{old}$), an inner product (which was the $L^2$ inner product in the $(t,x)$ coordinate system), and finally, a smooth topological structure for the theory's kinematically allowed states. Note that since the externalization map, \mbox{$J:S_\text{neutral}^\text{kin}\to S_\text{new}^\text{all}$}, is injective we have a bijection, $J:S_\text{new}^\text{kin}\leftrightarrow S_\text{neutral}^\text{kin}$ between our two characterizations of the theory's kinematically allowed states. We can use this bijection to transfer over all of the old theory's dynamically and metaphysically relevant structure onto $S_\text{new}^\text{kin}$ as follows:
\begin{align}
J:V_\text{neutral}^\text{kin}\to V_\text{new}^\text{kin}\coloneqq (S_\text{new}^\text{kin},\mathcal{T}_\text{new},+_\text{new},\,\cdot_\text{new}\,,\,\langle,\rangle_\text{new}).
\end{align}
Note that these new structures on $S_\text{new}^\text{kin}$ are exactly those induced by the $J$ map (i.e., they are defined analogously to Eq.~\eqref{InducedPlusTimes}).

But what are these induced structures exactly? It follows from familiar properties of the Fourier transform that $+_\text{new}$ and $\,\cdot_\text{new}\,$ are the point-wise addition and scalar multiplications operations on $\mathcal{M}_\text{new}$. Moreover, it follows that $\,\langle,\rangle_\text{new}$ is the $L^2$ inner product in the new theory's $(\tau,q)$ coordinate system. Thus, given our above choice of $J$ the new theory's induced structures all seem to fit very well on its new spacetime manifold. It should be stressed, however, that this is not guaranteed to happen in general. For instance, a different choice of $J$ might induce  a different inner product, $\,\langle,\rangle_\text{new}$, than the standard $L^2$ inner product. Alternatively, suppose that (like the Quartic Klein-Gordon theory) the non-local QHO theory had come equipped with a pointwise product operation, $\times_\text{old}$, on $\mathcal{M}_\text{old}$. In this case, the induced product operation in the new theory, $\times_\text{new}$, would have been a convolutional product on $\mathcal{M}_\text{new}$. It is important to note that the new theory's induced structures follow solely from our choice of injective map, $J:S_\text{neutral}^\text{kin}\to S_\text{new}^\text{all}$. They are not arrived at by identifying which operations are natural (e.g., pointwise or local) on $\mathcal{M}_\text{new}$. Although, of course, our choice of PSTOs might be guided by whether or not a nice $J$ map will be available to us.

\subsection{But are we forced to have $J$ implement a Fourier transform?}
The remainder of this section will discuss in more generality the latter half of the externalization process (i.e., picking an injector \mbox{$J:S_\text{neutral}^\text{kin}\to S_\text{new}^\text{all}$} given some PSTOs). The following questions will be answered in turn: What are the minimal constraints must our choice of $J$ satisfy? For a generic set of PSTOs, how do we know that such a $J$ map exists? Given that such a $J$ map exists, to what extent is it uniquely determined by our PSTOs?

Let us begin by placing some constraints on our choice of $J$. Recall that our goal is to build a new spacetime setting for the theory in question such that every pre-spacetime translation operation, $g\in G_\text{trans}$, becomes an honest-to-goodness spacetime translation in the new theory. We have so far seen how externalization generates a diffeomorphism $H(g)\in\text{Diff}(\mathcal{M}_\text{new})$ from every $g\in G_\text{trans}$. In order for this to truly count as a spacetime translation, however, it needs to act in the right way on the new theory's states, $\varphi_\text{new}=J(\bm{\varphi})$. Establishing this will require that we demand some compatibility between our choice of injector $J$ and our choice of PSTOs.

For any Lie group action $g\in G_\text{trans}$, we already know how it acts on any spacetime-neutral state, \mbox{$\bm{\varphi}\in S_\text{neutral}^\text{kin}$}. Namely, it acts as $\bm{\varphi}\mapsto g\,\bm{\varphi}$. In addition to converting every $g\in G_\text{trans}$ into a diffeomorphism $H(g)\in H_\text{trans}\subset\text{Diff}(\mathcal{M}_\text{new})$, externalization also converts every spacetime-neutral states, $\bm{\varphi}\in S_\text{neutral}^\text{kin}$ into a state on the new spacetime,  $\varphi_\text{new}=J(\bm{\varphi})\in S_\text{new}^\text{kin}$. Our choice of $J$ must be such that these two externalization maps, $g\mapsto H(g)$ and $\bm{\varphi}\mapsto J(\bm{\varphi})$, are compatible:
\begin{align}\label{EqEgh*E}
\forall g \ \forall \bm{\varphi} \ \ (J\circ g)\,\bm{\varphi} &= (H(g)^*\circ J)\,\bm{\varphi}.
\end{align}
with $g$ and $\bm{\varphi}$ ranging over  $G_\text{trans}$ and $S_\text{neutral}^\text{kin}$ respectively. Demanding Eq.~\eqref{EqEgh*E} of our injector $J$ ensures that externalization preserves not only the Lie group structure of $G_\text{trans}$, but also its structure as a Lie group action.

We must, however, demand slightly more than Eq.~\eqref{EqEgh*E} in order to ensure our desired result, i.e., that the action of $g\in G_\text{trans}$ on $S_\text{neutral}^\text{kin}$ is faithfully represented by the action of $H(g)^*\in H_\text{trans}{}^*$ on $S_\text{neutral}^\text{kin}$. Given only Eq.~\eqref{EqEgh*E} we could conceivably presently have distinct $g_1\neq g_2$ which correspond to identical actions on $S_\text{neutral}^\text{kin}$, namely $H(g_1)^*\vert_\text{kin}=H(g_2)^*\vert_\text{kin}$. If this were to happen, then $H(g)^*$'s action would not be a faithful representation of the $g$'s action. In order to avoid this we must demand that for all diffeomorphisms, $h\in H_\text{trans}\subset\text{Diff}(\mathcal{M}_\text{new})$, we have,
\begin{align}\label{EqhJphiJphi}
&\Big(\forall \bm{\varphi} \ \ h^* J(\bm{\varphi})=J(\bm{\varphi})\Big) \ \ \Longleftrightarrow \ \ h=\openone,
\end{align}
with $\bm{\varphi}$ ranging over $S_\text{neutral}^\text{kin}$. One can easily check that our above-discussed choice of $J$ (i.e., the one which implements a Fourier transform) satisfies these two conditions.

To review, externalization builds a new spacetime setting for our theory from the following materials: A pair of Lie groups, $G_\text{trans}$ and $G_\text{fix}$, acting on $\mathcal{S}_\text{neutral}^\text{kin}$ with faithful construction compatibility; a value space, $\mathcal{V}_\text{new}$, for the new theory; and an injective map, \mbox{$J:\mathcal{S}_\text{neutral}^\text{kin}\to S_\text{new}^\text{all}$}. This construction is considered successful if Eq.~\eqref{EqEgh*E} and Eq.~\eqref{EqhJphiJphi} are satisfied. But given some pair of Lie group actions, $G_\text{trans}$ and $G_\text{fix}$, how do we know that such value spaces, $\mathcal{V}_\text{new}$, and injective maps, \mbox{$J$}, exist? Their existence is guaranteed by demanding that $G_\text{trans}$ and $G_\text{fix}$ have injection compatibility.
\begin{quote}
{\hypertarget{InjCompat1}{\bf Definition: Injection Compatibility}}\footnote{A keen reader may note that what is demanded in Eq.~\eqref{JInjectionEq2} is slight more than what is needed to satisfy Eq.~\eqref{EqhJphiJphi}. The former quantifies over more diffeomorphisms than the latter does. As~\cite{ISEEquiv} discusses, making this only very slightly stronger demand greatly simplifies our analysis of the ISE Methodology.} - Let $G_1$ be a Lie group which acts smoothly on some space $\mathcal{S}_\text{N}=(S_\text{N},\mathcal{T}_\text{N},\dots)$. Let $G_2\subset G_1$ be a closed subgroup thereof. Let $\mathcal{M}_\text{G}\coloneqq G_1/G_2$ be the quotient manifold of $G_2$ and $G_1$. For any value space, $\mathcal{V}_\text{G}$, let $S_\text{G}^\text{all}$ be the space of all $\mathcal{V}_\text{G}$-valued functions definable on $\mathcal{M}_\text{G}$.

This pair of Lie group actions, $G_2\subset G_1$, on $\mathcal{S}_\text{N}$ have \textit{injection compatibility} if and only if the following condition holds. There exists a value space, $\mathcal{V}_\text{G}$, and an injective map, \mbox{$J_\text{G}:S_\text{N}\to S_\text{G}^\text{all}$}, with the following two properties: Firstly, we must have,
\begin{align}\label{JInjectionEq1}
&\forall g \ \forall \bm{\sigma} \ \ \Big((J_\text{G}\circ g)\,\bm{\sigma} = (\bar{g}^*\circ J_\text{G})\,\bm{\sigma}\Big),
\end{align}
and, secondly, for all diffeomorphisms, $d\in \text{Diff}(\mathcal{M}_\text{G})$, we must have,
\begin{align}\label{JInjectionEq2}
&\Big(\forall \bm{\sigma} \ \ d^* J_\text{G}(\bm{\sigma})=J_\text{G}(\bm{\sigma})\Big) \ \ \Longleftrightarrow \ \ d=\openone.
\end{align}
In both of these expressions $g$ ranges over $G_1$ and $\bm{\sigma}$ ranges over $\mathcal{S}_\text{N}$.
\end{quote}
Together with faithful construction compatibility, we are now demanding the following from our PSTOs. From $G_\text{trans}$ and $G_\text{fix}$ we can build a quotient manifold $\mathcal{M}_\text{G}=G_\text{trans}/G_\text{fix}$ such that: 1) the acted-upon space, $S_\text{neutral}^\text{kin}$, can be nicely mapped onto scalar field states on $\mathcal{M}_\text{G}$; and 2) $G_\text{trans}$ is faithfully represented as diffeomorphisms on $\mathcal{M}_\text{G}$  \textit{which act on} these states in the right way. This justifies the following gloss of the externalization process: Externalization builds a new spacetime by taking some hand-picked pre-spacetime translation operations, $G_\text{fix}\subset G_\text{trans}$ acting on $S_\text{neutral}^\text{kin}$, and letting them be honest-to-goodness spacetime translations, $H_\text{fix}\subset H_\text{trans}\subset\text{Diff}(\mathcal{M}_\text{new})$, acting on $S_\text{new}^\text{kin}$.

Now that we have identified the general conditions which our choice injective map \mbox{$J:S_\text{neutral}^\text{kin}\to S_\text{new}^\text{all}$} must satisfy (namely, Eq.~\eqref{EqEgh*E} and Eq.~\eqref{EqhJphiJphi}), we are in a position to ask: Were any other options for $J$ were available to us in externalizing the non-local QHO theory? Recall that prior to our choice of $J$ we had already constructed $\mathcal{M}_\text{new}$ from our Fourier-shift PSTOs and noted its close relationship with the old theory's Fourier space. Hence, we are really asking the following: Is there any way to populate the new theory's spacetime (i.e., a copy of the old theory's Fourier space) other than with the Fourier transform of the old theory's states? 

The remainder of this section will be spent showing that (up to a metaphysically irrelevant rescaling freedom) the above-discussed constraints actually force the intuitive choice of $J$ upon us. To see this, let us take
\begin{align}
\varphi_\text{old}^{(\lambda_1,\lambda_2)}(t,x)\coloneqq\exp(\ii\,\lambda_1\,t) \, \delta(x-\lambda_2)
\end{align}
and note that the Fourier transform of this state, 
\begin{align}
\tilde{\varphi}_\text{old}^{(\lambda_1,\lambda_2)}(t,k)=\exp(\ii\,\lambda_1\,t)\exp(\ii\,\lambda_2\,k)
\end{align}
is a planewave \textit{in the old theory's Fourier space}. Namely, $\tilde{\varphi}_\text{old}^{(\lambda_1,\lambda_2)}(t,k)$ is a simultaneous eigenvector of both $\partial_t$ and $\partial_k$ with eigenvalues $\ii\lambda_1$ and $\ii\lambda_2$ respectively. These differential operators are, of course, the generators of translations in the old theory's Fourier space; We have $f(\mathrm{d}t,0)=\openone+\partial_t$ and $f(0,\mathrm{d}k)=\openone+\partial_k$. Note that $\varphi_\text{old}^{(\lambda_1,\lambda_2)}(t,x)$ is a simultaneous eigenvector of $f(\mathrm{d}t,0)$ and $f(0,\mathrm{d}k)$ with eigenvalues $1+\ii\lambda_1$ and $1+\ii\lambda_2$ respectively.

Per the above discussion, for every Fourier shift in the old theory, $f(\Delta t,\Delta k)$, there is a corresponding diffeomorphism on $\mathcal{M}_\text{new}$, namely $h(\Delta t,\Delta k)$. In particular, corresponding to the infinitesimal Fourier shifts, $f(\mathrm{d}t,0)$ and $f(0,\mathrm{d}k)$, there are corresponding infinitesimal diffeomorphisms on $\mathcal{M}_\text{new}$, namely $h(\mathrm{d}t,0)$ and $h(0,\mathrm{d}k)$. Lifting these diffeomorphisms in the new spacetime setting we have  $h(\mathrm{d}t,0)^*=\openone+\partial_\tau$ and $h(0,\mathrm{d}k)^*=\openone+\partial_q$. Eq.~\eqref{EqEgh*E} demands that these lifted diffeomorphisms act on the new theory's states in the same way that the above-discussed Fourier shifts acted on the old theory's states. In particular, since $\varphi_\text{old}^{(\lambda_1,\lambda_2)}$ is a simultaneous eigenvector of $f(\mathrm{d}t,0)$ and $f(0,\mathrm{d}k)$, it follows that the new state,
\begin{align}
\varphi_\text{new}^{(\lambda_1,\lambda_2)} \coloneqq J\circ\mathcal{F}_\text{vec} (\varphi_\text{old}^{(\lambda_1,\lambda_2)}),
\end{align}
must be a simultaneous eigenvector of $h(\mathrm{d}t,0)^*$ and $h(0,\mathrm{d}k)^*$. Concretely, we have,
\begin{align}
\partial_\tau\,\varphi_\text{new}^{(\lambda_1,\lambda_2)}
=\ii\lambda_1\, \varphi_\text{new}^{(\lambda_1,\lambda_2)}\quad\text{and}\quad \partial_q\,\varphi_\text{new}^{(\lambda_1,\lambda_2)}
=\ii\lambda_2\, \varphi_\text{new}^{(\lambda_1,\lambda_2)}
\end{align}
That is, $\varphi_\text{new}^{(\lambda_1,\lambda_2)}$ must be a simultaneous eigenvector of $\partial_\tau$ and $\partial_q$ with eigenvalues $\ii\lambda_1$ and $\ii\lambda_2$. Namely,  $\varphi_\text{new}^{(\lambda_1,\lambda_2)}$ must be a planewave in the new theory's spacetime.

But what does this tell us about which injective maps $J$ are allowed? Recall that $\varphi_\text{old}^{(\lambda_1,\lambda_2)}(t,x)$ is a planewave \textit{in the old theory's Fourier space}. We thus have a planewave-to-planewave correspondence between the old theory's Fourier space and the new theory's spacetime. Concretely, Eq.~\eqref{EqEgh*E} enforces that we have
\begin{align}
\varphi_\text{new}^{(\lambda_1,\lambda_2)}(\tau,q) &= a(\lambda_1,\lambda_2) \, \exp(\ii\,\lambda_1\,\tau) \, \exp(\ii\,\lambda_2\,q)\\
\nonumber
&=a(\lambda_1,\lambda_2) \ \tilde{\varphi}_\text{old}^{(\lambda_1,\lambda_2)}(\tau,q)
\end{align}
for some scaling parameter $a(\lambda_1,\lambda_2)$. A quick calculation shows that a generic $\varphi_\text{old}(t,x)\in S_\text{old}^\text{kin}$ (which has Fourier transform, $\tilde{\varphi}_\text{old}(t,k)$) must be mapped to
\begin{align}
J\circ\mathcal{F}_\text{vec}:\varphi_\text{old}\mapsto
\varphi_\text{new}(\tau,q)&= A(\tau,q) *_\text{conv} \tilde{\varphi}_\text{old}(\tau,q)
\end{align}
for some profile $A(\tau,q)$. Recalling that convolution in space corresponds to point-wise multiplication in Fourier space, this $A(\tau,q)$ freedom captures our ability to freely rescale each of the new theory's Fourier modes via the above-discussed $a(\lambda_1,\lambda_2)$. Note that our intuitive choice of $J$ (namely, Eq.~\eqref{IntuitiveJFourier}) corresponds to picking a scaling of $a(\lambda_1,\lambda_2)=1$ for each of the new theory's planewaves and hence a profile $A(\tau,q)=1$. Up to this rescaling freedom, there is only one way to map the old theory's states onto the new theory's spacetime (namely, via a Fourier transform).

But what should we make of this rescaling freedom? I will now argue that this planewave-rescaling freedom is metaphysically irrelevant and that it is merely a superficial aspect of how we are representing the theory's states in the new spacetime setting. To see this, recall that the non-local QHO theory came equipped with a metaphysically relevant inner product, $\langle\varphi_\text{old},\phi_\text{old}\rangle_\text{old}$, for comparing the sizes and similarity of states. In particular, this was the $L^2$ product in the old theory's $(t,x)$ coordinate system. Note that according to this inner product, the old theory's planewave states, $\varphi_\text{old}^{(\lambda_1,\lambda_2)}(t,x)$, are normalized (to a Dirac delta). 

Post-internalization a metaphysically relevant inner product, $\langle\bm{\varphi},\bm{\phi}\rangle_\text{neutral}$ is induced on $V_\text{neutral}^\text{kin}$. Post-externalization a metaphysically relevant inner product, $\langle\varphi_\text{new},\phi_\text{new}\rangle_\text{new}$ is induced on $V_\text{new}^\text{kin}$. Perhaps surprisingly, according to this new inner product, the new planewave states $\varphi_\text{new}^{(\lambda_1,\lambda_2)}(\tau,q)$ are normalized no matter what $a(\lambda_1,\lambda_2)$ we choose. Thus, according to the metaphysically relevant way of judging the relative size and similarity of the new theory's states our planewave-rescaling freedom simply does not matter. 

As the above discussion has shown, the externalization process is driven almost entirely by our choice of PSTOs. These determine not only the new theory's spacetime manifold (up to diffeomorphism) but also how the old theory's states are to be mapped onto this new spacetime setting (up to some metaphysically irrelevant rescaling of the new theory's Fourier modes).

\section{Example Application: Moving from a Lattice Theory to a Continuous Spacetime Theory}\label{SecDemoISE}
This section will show how one can use the ISE Methodology to redescribe a lattice theory (i.e., a theory set on a discrete spacetime, $\mathcal{M}_\text{old}\cong\mathbb{R}\times\mathbb{Z}$) as existing on a continuous spacetime manifold, $\mathcal{M}_\text{new}\cong\mathbb{R}\times\mathbb{R}$. Before this, however, some mathematical definitions are needed relating to various discrete approximations of the derivative. 

Consider the nearest-neighbor and next-to-nearest-neighbor approximations of the second derivative,
\begin{align}
\Delta_{(1)}^2/\epsilon^2:\ \partial_x^2 f(x)
&\approx\frac{f(x+\epsilon)-2 f(x)+f(x-\epsilon)}{\epsilon^2}\\
\nonumber
\Delta_{(2)}^2/\epsilon^2:\ \partial_x^2 f(x)
&\approx\frac{-f(x+2\epsilon)+16\,f(x+\epsilon)-30\, f(x)+16\,f(x-\epsilon)-f(x+2\epsilon)}{12\,\epsilon^2}.
\end{align}
Closely, associated with these derivative approximations are the following two infinite matrices,
\begin{align}\label{BigToeplitz}
\Delta_{(1)}^2\coloneqq\text{Toeplitz}(&\dots,0,0,0,0,1,{\bf-2},1,0,0,0,0,\dots)\\
\nonumber
\Delta_{(2)}^2\coloneqq\text{Toeplitz}(&\dots,0,0,0,\frac{-1}{12},\frac{16}{12},{\bf \frac{-30}{12}},\frac{16}{12},\frac{-1}{12},0,0,0,\dots).
\end{align}
Toeplitz matrices are the so-called diagonal-constant matrices with \mbox{$[A]_{i,j}=[A]_{i+1,j+1}$}. Thus, the values listed in the above expression give the matrix's values on either side of the main diagonal. The value in the middle of the list (in bold) corresponds to the main diagonal. 

These two infinite matrices can be easily generalized to, $\Delta_{(n)}^2$, the Toeplitz matrix related to the $\text{(next-to)}^n$-nearest-neighbor approximation of the second derivative. Taking the limit $n\to\infty$ one can define,
\begin{align}
D^2\coloneqq\text{Toeplitz}(&\dots,\frac{2}{25},\frac{-2}{16},\frac{2}{9},\frac{-2}{4},\frac{2}{1},{\bf \frac{-2\pi^2}{6}},\frac{2}{1},\frac{-2}{4},\frac{2}{9},\frac{-2}{16},\frac{2}{25},\dots),
\end{align}
as the infinite Toeplitz matrix related to the infinite-range discrete approximation of the second derivative. One can similarly define an infinite Toeplitz matrix, $D$, related to the infinite-range discrete approximation of the first derivative,
\begin{align}\label{Ddef}
D\coloneqq\text{Toeplitz}(&\dots,\frac{-1}{5},\frac{1}{4},\frac{-1}{3},\frac{1}{2},-1,{\bf 0},1,\frac{-1}{2},\frac{1}{3},\frac{-1}{4},\frac{1}{5},\dots).
\end{align}
As it turns out, $D^2$ is the square of $D$. 

For later reference, it should be noted how the above four Toeplitz matrices act on the discrete planewaves, $\varphi_\text{old}^{(\omega,k)}(t,n)\coloneqq\exp(\ii \omega t + \ii k\,n)$, with $\omega\in\mathbb{R}$ and $k\in[-\pi,\pi]$,\footnote{Note that these discrete planewaves are period in their wavenumber $k$ with a period of $2\pi$. Namely, it follows from Euler's identity that $\varphi_\text{old}^{(\omega,k+2\pi)}$ and $\varphi_\text{old}^{(\omega,k)}$ are literally the same functions of $(t,n)$.}
\begin{align}\label{D2DEigenvalues}
\Delta_{(1)}^2 \, \varphi_\text{old}^{(\omega,k)} &= (2\cos(k)-2) \ \varphi_\text{old}^{(\omega,k)}\\
\nonumber
\Delta_{(2)}^2 \, \varphi_\text{old}^{(\omega,k)} &= \frac{1}{12}(-2\cos(2k) +32\cos(k)-30) \ \varphi_\text{old}^{(\omega,k)}\\
\nonumber
D^2 \, \varphi_\text{old}^{(\omega,k)} &= -k^2 \  \varphi_\text{old}^{(\omega,k)}\\
\nonumber
D \, \varphi_\text{old}^{(\omega,k)} &= \ii k \  \varphi_\text{old}^{(\omega,k)}.
\end{align}
Notably, the discrete planewave, $\varphi_\text{old}^{(\omega,k)}$, is an eigenvector of each of these linear operators with different eigenvalues. Incredibly, the eigenvalues for $D^2$ and $D$ are $-k^2$ and $\ii k$ exactly matching the eigenvalues of the continuum derivatives, $\partial_x^2$ and $\partial_x$.

Given these definitions, we can now state the following three lattice theories:\footnote{In these theories, space is discrete whereas time is continuous. The ISE Methodology can also be applied when both space and time are discrete. For instance, in ~\cite{DiscreteGenCovPart2} these techniques are applied to a discrete version of the Klein-Gordon equation.}
\begin{quote}
\hypertarget{MDS}{{\bf Heat Equation on a Discrete Spacetime}} - Consider a spacetime theory about a scalar field, $\varphi_\text{old}:\mathcal{M}_\text{old}\to\mathcal{V}_\text{old}$, with spacetime, $\mathcal{M}_\text{old}\cong\mathbb{R}\times\mathbb{Z}$, and value space, $\mathcal{V}_\text{old}\cong\mathbb{R}$. This theory's states are subject to the following kinematic constraint: In some fixed global coordinate system, $(t,n)$, each field must be smooth. In this coordinate system, the metaphysically relevant way to judge the relative size and similarity of this theory's states is with an $L^2$ inner product. There are three variants of this theory obeying the following three dynamical equations respectively: 
\begin{align}
{\bf H1:}& \quad \frac{\d }{\d t}\vec{\varphi}_\text{old}(t)
=\alpha\,\Delta_{(1)}^2\, \vec{\varphi}_\text{old}(t)\\
\nonumber
{\bf H2:}& \quad \frac{\d }{\d t}\vec{\varphi}_\text{old}(t)
=\alpha\,\Delta_{(2)}^2\, \vec{\varphi}_\text{old}(t)\\
\nonumber
{\bf H3:}& \quad \frac{\d }{\d t}\vec{\varphi}_\text{old}(t)
=\alpha\,D^2\, \vec{\varphi}_\text{old}(t)
\end{align}
for some $\alpha>0$.
\end{quote}
These are the nearest neighbor (H1), next-to-nearest neighbor (H2), and infinite-range (H3) discrete heat equations respectively.

For brevity, the internalization of this theory must be skipped over. It proceeds along very similar grounds to the internalization of the non-local QHO theory discussed at the end of Sec.~\ref{SecSurInt}. After internalization, we next search for good PSTOs to externalize.  Perhaps surprisingly, these lattice theories also admit continuous pre-spacetime translation operations. In particular, these are the smooth transformations sketched in the third column of Fig.~\ref{FigPSTOs}. As this example will show, by externalizing these PSTOs we can find new continuous spacetime settings for our formerly discrete theories. 

The applicability of (an early version of) the ISE Methodology to these discrete spacetime theories was first demonstrated in \cite{DiscreteGenCovPart1,DiscreteGenCovPart2}. What follows is a condensed and updated version of this treatment. My analysis of these discrete spacetime theories is largely inspired by the work of the mathematical physicist Achim Kempf.\footnote{For an overview of Kempf's work on these topics see~\cite{Kempf2018}.} See, for instance, \cite{Kempf_2010} entitled, ``Spacetime could be simultaneously continuous and discrete, in the same way that information can be'' where such a continuous-to-discrete correspondence is understood using the Nyquist-Shannon sampling theory. 

But how is it that the above-discussed lattice theories can have both continuous and discrete spacetime representations? Consider first the discrete shift diffeomorphism, \mbox{$d_{+m}:(t,n)\mapsto (t,n+m)$} for $m\in\mathbb{Z}$, acting on the old theory's spacetime manifold, $\mathcal{M}_\text{old}$. Due to the discreteness of the old spacetime manifold, we cannot promote $m\in\mathbb{Z}$ to a continuous parameter, $\epsilon\in\mathbb{R}$. Next, consider lifting these discrete shift maps to act on $\varphi_\text{old}\in V_\text{old}^\text{kin}$ as $d_{+m}^*:V_\text{old}^\text{kin}\to V_\text{old}^\text{kin}$. Unlike the old spacetime, $\mathcal{M}_\text{old}$, the state space, $V_\text{old}^\text{kin}$, is continuous; Hence, if we focus on the lifted maps, $d_{+m}^*$, we might be able to promote $m\in\mathbb{Z}$ to a continuous parameter, $\epsilon\in\mathbb{R}$. This is how we will find our continuous PSTOs. 

Recall that $D$ (defined in Eq.~\eqref{Ddef}) is the best infinite-range discrete approximation of the derivative. Consider the smooth transformations on $V_\text{old}^\text{kin}$ given by  $\text{exp}(-\epsilon\, D)$ for $\epsilon\in\mathbb{R}$. It turns out that for $\epsilon=m\in\mathbb{Z}$ we have $\text{exp}(-\epsilon\, D)=d_{+m}^*$. That is, the continuous family of smooth transformation, $\text{exp}(-\epsilon D)$, nicely interpolates between the discrete shift operations. It is the smooth action of these $\text{exp}(-\epsilon D)$ maps which is shown in the third column of Fig.~\ref{FigPSTOs}.

Let us now construct a set of continuous PSTOs. Let the maps, 
\begin{align}\label{DefFCont}
f(\Delta t,\epsilon) \, \vec{\varphi}_\text{old}(t)\coloneqq\text{exp}(-\epsilon D) \, \vec{\varphi}_\text{old}(t-\Delta t),    
\end{align}
form a Lie group $f(\Delta t,\epsilon)\in F_\text{Cont}\cong(\mathbb{R}^2,+)$ which acts on $V_\text{old}^\text{kin}$. These transformations faithfully survive internalization becoming a Lie group $g(\Delta t,\epsilon)\in G_\text{Cont}$ acting on $V_\text{neutral}^\text{kin}$. Let us take $G_\text{trans}^{\text{(ext)}}=G_\text{Cont}\cong(\mathbb{R}^2,+)$ to be our PSTOs. As in the Fourier example, it follows from $G_\text{trans}^{\text{(ext)}}$ being abelian that we must pick $G_\text{fix}^{\text{(ext)}}=\{\openone\}$.

Constructing a new spacetime manifold from these PSTOs we have
\begin{align}
\mathcal{M}_\text{new}\cong G_\text{trans}^{\text{(ext)}}/G_\text{fix}^{\text{(ext)}}\cong\mathbb{R}^2.    
\end{align}
By construction, the Lie group in the numerator, $g(\Delta t,\epsilon)\in G_\text{Cont}$, will be faithfully represented as a diffeomorphism, $h(\Delta t,\epsilon)\coloneqq H(g(\Delta t,\epsilon))\in \text{Diff}(\mathcal{M}_\text{new})$, on the new spacetime. As in the previous examples, it will be helpful to introduce a coordinate system associated with these diffeomorphisms. Let $(\tau,q)$ be a coordinate system for $\mathcal{M}_\text{new}\cong\mathbb{R}^2$ such that such that $h(\Delta t,\epsilon)(\tau,q)=(\tau+\Delta t,q+a\,\epsilon)$. Note that I have defined the $(\tau,q)$-coordinates such that a unit shift of $\epsilon$ in the old spacetime corresponds to a shift of the $q$-coordinate by a distance $a$ in the new theory. Externalization has turned our choice of $G_\text{trans}=G_\text{Cont}$ into honest-to-goodness spacetime translations on $\mathcal{M}_\text{new}$.

We next need to map the old theory's states and dynamics onto this new spacetime setting by picking out an injective map,
\mbox{$J:S_\text{neutral}^\text{kin}\to S_\text{new}^\text{all}$}, which obeys Eq.~\eqref{EqEgh*E} and Eq.~\eqref{EqhJphiJphi} are satisfied. Namely, the new diffeomorphisms,
$h(\Delta t,\epsilon)\in \text{Diff}(\mathcal{M}_\text{new})$, must lift to act in the right way on the new theory's states, $\varphi_\text{new}=J(\bm{\varphi})$. As in the Fourier example, an eigenvector argument will fix our choice of $J$ up to a metaphysically irrelevant rescaling of the new theory's Fourier modes.

In particular, it follows from Eq.~\eqref{EqEgh*E} that infinitesimal action by $f(\Delta t,\epsilon)$ in the old theory must correspond to infinitesimal action by $h(\Delta t,\epsilon)$ in the new theory. It follows from this that a generic discrete planewave in the old theory's spacetime, must map onto a corresponding continuous planewave in the new theory's spacetime. To see this, consider a generic discrete planewave in the old theory's spacetime,
\begin{align}
\varphi_\text{old}^{(\omega,k)}(t,n)\coloneqq\exp(\ii \omega t + \ii k\,n)
\end{align} 
Note that this state is a simultaneous eigenvector of $\partial_t$ and $D$ with eigenvalues $\ii\omega$ and $\ii k$ respectively. It follows from Eq.~\eqref{EqEgh*E} that the corresponding state in the new theory, $\varphi_\text{new}^{(\omega\, k)}(\tau,q)\coloneqq J\circ\mathcal{F}_\text{vec}(\varphi_\text{old}^{(\omega\, k)})$, must be a simultaneous eigenvector of $\partial_\tau$ and $\partial_q$ with eigenvalues $\ii\omega$ and $\ii k/a$ respectively. Namely, we must have,
\begin{align}
J\circ\mathcal{F}_\text{vec}:\varphi_\text{old}^{(\omega\, k)}\mapsto\varphi_\text{new}^{(\omega\, k)}(\tau,q)= b(\omega,k) \ \text{exp}(\ii\omega \tau+\ii k q/a),
\end{align}
for some scaling $b(\omega,k)$. Thus, up to some rescaling freedom, the externalization process must map discrete planewaves in the old theory's spacetime onto a corresponding continuous planewave in the new theory's spacetime. Note that the old theory's discrete planewaves (which are exhausted by the wavenumbers $k\in[-\pi,\pi]$ and $\omega\in\mathbb{R}$) must become continuous planewaves with wavenumbers $k/a\eqqcolon\kappa\in[-\pi/a,\pi/a]$. Hence, all of the new theory's states are guaranteed to be \textit{bandlimited} with a bandwidth of $K\coloneqq\pi/a$: Namely, their support in Fourier space is limited to $\kappa\in[-K,K]$. Note that this conclusion holds independent of our choice of scalings, $b(\omega,k)$.

A quick calculation shows that a generic state, $\varphi_\text{old}(t,n)\in V_\text{old}^\text{kin}$, must be mapped onto
\begin{align}
J\circ\mathcal{F}_\text{vec}:\varphi_\text{old}\mapsto
\varphi_\text{new}(\tau,q)&= B(\tau,q/a) *_\text{conv} \sum_{n\in\mathbb{Z}} \delta(q/a-n) \ \varphi_\text{old}(\tau,n)
\end{align}
for some profile $B(\tau,q/a)$ given by the Fourier transform of $b(\omega,k)$ extended to vanish for $k\notin[-\pi,\pi]$. This bandlimited profile is convolved against a direct embedding of the discrete field state, $\varphi_\text{old}(\tau,n)$, as Dirac deltas onto the new continuous spacetime. 

Just as in the Fourier example, there is effectively only one way to map the old theory's states onto the new theory's spacetime. The only freedom we have in populating the new spacetime is in a free choice of scalings for each of the new theory's planewaves, $b(\omega,k)$ for $\omega\in\mathbb{R}$ and $k\in[-\pi,\pi]$. But what should we make of this planewave-rescaling freedom? Just as in the Fourier example, this rescaling freedom is metaphysically irrelevant if we judge the relative size and similarity of states using the new theory's induced inner product, $\langle,\rangle_\text{new}$. The only difference this choice makes is in how the theory's states are being represented in this new spacetime setting. Hence, we are free to pick $b(\omega,k)=1$ for all $\omega\in\mathbb{R}$ and all $k\in[-\pi,\pi]$. Noting that $b(\omega,k)=0$ outside of this region we have,
\begin{align}
J\circ\mathcal{F}_\text{vec}:\varphi_\text{old}\mapsto
\varphi_\text{new}(\tau,q)&= \sum_{n\in\mathbb{Z}} \text{sinc}(q/a-n) \ \varphi_\text{old}(\tau,n)
\end{align}
where $\text{sinc}(z)\coloneqq\sin(\pi\,z)/(\pi\,z)$.

Continuing with this choice of $J$ one can map the old theory's dynamics into this new spacetime setting in order to define the following theories:\footnote{See \cite{DiscreteGenCovPart1} for details.}
\begin{quote}
\hypertarget{BMCS}{{\bf Bandlimited Heat Equation on a Continuous Spacetime -}} Consider a spacetime theory about a scalar field, $\varphi_\text{new}:\mathcal{M}_\text{new}\to\mathcal{V}_\text{new}$, with spacetime, $\mathcal{M}_\text{new}\cong\mathbb{R}^2$, and value space, $\mathcal{V}_\text{new}\cong\mathbb{R}$. This theory's states are subject to the following kinematic constraint: In some fixed global coordinate system, $(\tau,q)$, each field must be smooth in the $\tau$ direction. Moreover, at each $\tau$-coordinate the field must have a well-defined Fourier transform, $\tilde{\varphi}_\text{new}(t,\kappa)$, with support only over $\kappa\in[-\pi/a,\pi/a]$. In this coordinate system, the metaphysically relevant way to judge the relative size and similarity of this theory's states is with an $L^2$ inner product. There are three variants of this theory obeying the following three dynamical equations respectively:
\begin{align}
{\bf H1:}&\quad\partial_\tau\varphi_\text{new}(\tau,q)
=\alpha \, [2\text{cosh}(a\partial_q)-2] \ \varphi_\text{new}(\tau,q)\\
\nonumber
{\bf H2:}&\quad\partial_\tau\varphi_\text{new}(\tau,q)=\frac{\alpha}{12} [-2\text{cosh}(2a\partial_q)+32\text{cosh}(a\partial_q)-30]\varphi_\text{new}(\tau,q)\\
\nonumber
{\bf H3:}&\quad \partial_\tau\varphi_\text{new}(\tau,q)
=\frac{\alpha}{a^2}\,\partial_q^2 \, \varphi_\text{new}(\tau,q)
\end{align}
for some $\alpha>0$.
\end{quote}
Note that H3 is the standard continuum heat equation (albeit bandlimited). H1 and H2 are similar to the continuum heat equation, but feature some non-locality at a length scale of order $\sim a$. 

One remarkable feature of these bandlimited heat equations should be noted: They each have a continuous translation symmetry in their $q$-coordinate. In the old theory, this corresponds to the fact that action by $\exp(-\epsilon D)$ is a dynamical symmetry. To see that it maps solutions to solutions, simply note that each of 
$\Delta_{(1)}^2$, $\Delta_{(2)}^2$, $D^2$, and $D$ are all diagonal in the same basis (see Eq.~\eqref{D2DEigenvalues}). In hindsight, one can understand this application of the ISE Methodology as follows: In Eq.~\eqref{DefFCont}, we identified a nice set of dynamical symmetries, $f(\Delta t,\epsilon)\in F_\text{Cont}\cong(\mathbb{R}^2,+)$, in our old lattice-based theory. Despite not being spacetime translations these transformations on $V_\text{old}^\text{kin}$ are structurally indistinguishable from spacetime translations. Namely, they are PSTOs. The ISE Methodology allows us to build a new spacetime setting for our theory such that these PSTOs become honest-to-goodness spacetime translations. Indeed, since these PSTOs are dynamical symmetries, they turn into spacetime symmetries of the new theory.

\section{Conclusion and Outlook}\label{SecConclusion}
This paper has introduced the ISE Methodology and demonstrated how it can be applied to two example theories. To review, the ISE Methodology is a three-step process (Internalize, Search, Externalize). The first step, internalization, allows one to divorce a theory's dynamical and kinematical content from any assumed topological background structure (i.e., the spacetime manifold, $\mathcal{M}_\text{old}$). In the searching step, one then picks out a set of pre-spacetime translation operations (PSTOs). These are a certain kind of smooth Lie group action on the theory's states which are, in a certain sense, structurally indistinguishable from spacetime translations. An intuitive example explored in Sec.~\ref{SecDemoExt} are translations, not in spacetime, but rather in Fourier space. Finally, the externalization step then builds a new spacetime setting, $\mathcal{M}_\text{new}$, from these hand-picked pre-spacetime translation operations such that they become honest-to-goodness spacetime translations on $\mathcal{M}_\text{new}$. In the Fourier example, the new theory's spacetime is effectively a copy of the old theory's Fourier space. Correspondingly, the new theory's states and dynamics are naturally related to those of the old theory by a Fourier transform.

In Sec.~\ref{SecDemoISE}, the exact same topological redescription techniques were used to redescribe a lattice theory (i.e., a theory set on a discrete spacetime, $\mathcal{M}_\text{old}\cong\mathbb{R}\times\mathbb{Z}$) as existing on a continuous spacetime manifold, $\mathcal{M}_\text{new}\cong\mathbb{R}\times\mathbb{R}$. This topological redescription is facilitated by the fact that the discrete spacetime theory in question admits a set of continuous pre-spacetime translation operations (see the third column of Fig.~\ref{FigPSTOs}). What makes these PSTOs is that, when viewed as a Lie group action on the theory's states, these continuous transformations are structurally identical to spacetime translations. Namely, they are structurally identical to spacetime translations in exactly the same sense as the above-discussed Fourier shifts were. Hence, they too qualify as PSTOs. In addition to being PSTOs, these continuous transformations are also dynamical symmetries (i.e., they map solutions to solutions). The ISE Methodology allows us to redescribe this lattice theory in such a way that its hidden dynamical symmetry is re-expressed as a spacetime symmetry in the new theory.

But of what philosophical use are these topological redescription techniques? I have here claimed that the ISE Methodology is a powerful tool for discovering and negotiating between a wide range of spacetime settings for a given spacetime theory. Namely, as~\cite{ISEEquiv} proves, by using the ISE Methodology one can access effectively every possible topological redescription of a given theory's dynamical and kinematical content, limited only by a weak spacetime-kinematic compatibility condition. Arguably, this puts our newly discovered capacity for topological redescription on par with our existing capacity for coordinate change (and for re-axiomization of theories stated in first-order logic). 

If an analogy between these various types of redescription can be developed, it should become possible to defend an analogous interpretation of them. It is common to view coordinate systems as a representational artifact which we project onto the world in the process of codifying the dynamical behavior of matter. Similarly, on a broadly Humean view of the laws of nature (roughly, fundamental laws as best-axioms) our theory's laws are thought to reflect nothing metaphysically substantial in the world beyond them being one particularly nice way (among others) of codifying the dynamical behavior of matter. Aspirationally, the ISE Methodology could be used to support an analogous view of the spacetime manifold: The spacetime manifolds which appears in our best scientific theories reflect nothing metsphysically substantial in the world beyond them being one particularly nice way (among others) of \textit{codifying} the dynamical behavior of matter.

\subsection*{Acknowledgements}
The author thanks James Read, Tushar Menon, Nick Huggett, Harvey Brown, Oliver Pooley, Adam Caulton, and Neils Linneman for their helpful feedback.

\subsection*{Declarations}
The author did not receive support from any organization for the submitted work. The author has no competing interests to declare that are relevant to the content of this article.

%% file: ChapAndApp-Arxiv/A1_Homo.tex
\section{An Introduction to Smooth Homogeneous Manifolds}\label{AppHomoMan}
A smooth manifold, $\mathcal{M}$, is said to be homogeneous simpliciter if there exists a finite-dimensional Lie group, $G$, which it is homogeneous with respect to. A smooth manifold, $\mathcal{M}$, is said to be homogeneous with respect to a finite-dimensional Lie group, $G$, if there exists a group action, $\theta:G\times \mathcal{M}\to \mathcal{M}$, which acts smoothly and transitively on $\mathcal{M}$. A group action, $\theta$, is said to act transitively on $\mathcal{M}$ when for every pair of points, $p,q\in \mathcal{M}$, there is some $g\in G$ with $\theta(g,p)=q$. That is, a group action is transitive when it can map any point anywhere.

The intuition behind this definition of homogeneity is that all points on such a manifold, $\mathcal{M}$, are equivalent from the perspective of the Lie group, $G$: Concretely, any point $q\in \mathcal{M}$ can be replaced by any other point $p\in \mathcal{M}$ via action by some $g\in G$. Said differently, there is only one $G$-orbit, namely the whole manifold $\mathcal{M}$. If $G$'s action on $\mathcal{M}$ wasn't transitive, then it would divide $\mathcal{M}$ into several distinct pieces (i.e., the several $G$-orbits). Hence, if $G$'s action on $\mathcal{M}$ fails to be transitive, then from $G$'s perspective $\mathcal{M}$ would seem inhomogeneous (i.e., with several different kinds of points, namely $G$-orbits). 

For example, $\mathbb{R}^2$ is homogeneous under the group of rigid translations. Similarly, $S^2$ is homogeneous under the group of rigid rotations. It should be noted that while these two examples feature the geometrically loaded term ``rigid'', the sense in which these manifolds are homogeneous has nothing to do with their geometry. In particular, homogeneous does not mean constant curvature. Whether or not a smooth manifold is homogeneous is solely a matter of topology.

A smooth manifold can fail to be homogeneous on purely topological grounds. For a simple example consider $\mathcal{M}\cong \mathbb{T}^2\,\dot{\cup}\,\mathbb{R}^2$, the disjoint union of a torus and a two-dimensional plane. No smooth transformation on $\mathcal{M}$ can map a torus-point to a plane-point. For a less trivial example of a non-homogeneous manifold, consider the double torus, $\mathcal{M}\cong \mathbb{T}^2\#\mathbb{T}^2$. The non-homogeneity of this manifold hinges crucially upon us restricting our attention to finite-dimensional Lie groups.\footnote{All Lie groups which act smoothly and transitively over the double torus are infinite-dimensional. This follows from \cite{GMostow2005}'s proof that all smooth connected manifolds which support smoothly and transitively-acting finite-dimensional Lie groups actions have Euler characteristic, $\chi\geq0$. Note that the double torus has $\chi=-2$.} Namely, if we were to widen the definition of smooth homogeneous manifolds to allow for infinite-dimensional Lie groups, then the double torus would count as homogeneous. In fact, under this extended definition, a smooth manifold is homogeneous if and only if its connected components are pair-wise diffeomorphic. For any such manifold, the infinite-dimensional Lie group of diffeomorphisms, $\text{Diff}(\mathcal{M})$, acts smoothly and transitively over $\mathcal{M}$. If we make this allowance for infinite-dimensional Lie groups, then the only way in which a smooth manifold can be inhomogeneous is by having dissimilar connected components. Note that under this extended definition every smooth connected manifold would be homogeneous. For mathematical simplicity, however, this dissertation will use the more restrictive finite-dimensional definition of homogeneity.

The remainder of this appendix is organized as follows. 
Sec.~\ref{AppConChaTheorems} will present several definitions and theorems regarding smooth homogeneous manifolds which are of central importance to the ISE Methodology.\footnote{For a general introduction to Lie groups and homogeneous spaces, see \cite{IntroToLieGeoHomo} particularly Chapter 4. For more technical details relating to the basic theorems of differential topology invoked below, see \cite{WarnerFrankW1983Fodm,BrickellF1970Dm:a,KobayashiShoshichi1996Fodg}.} To overview, it is well known that the quotient of any two Lie groups is a smooth homogeneous manifold.\footnote{Technically, not any pair of Lie groups will do. The group in the denominator needs to be a closed subgroup of the Lie group in the numerator.} Moreover, it is well known that any homogeneous manifold is diffeomorphic to the quotient of some two Lie groups. Hence, up to diffeomorphism the smooth homogeneous manifolds are exactly these Lie group quotients. The theorems needed for the ISE Methodology, however, are slightly stronger. These improved theorems (stated below) show that, up to diffeomorphism, any smooth homogeneous manifold, $\mathcal{M}$, can be faithfully reconstructed as a quotient of \textit{a certain kind} of Lie groups. Namely, we can take the Lie groups to be among $\mathcal{M}$'s diffeomorphisms and, moreover, to have a certain kind of faithful construction compatibility (defined below). These slightly stronger theorems will be proved in Sec.~\ref{AppHomoManConThm} and Sec.~\ref{AppHomoManChaThm}. Following this, Sec.~\ref{AppSphereRecon} will give a concrete demonstration of these reconstruction techniques for the 2-sphere, $\mathcal{M}\cong S^2$.

\subsection{Constructing and Characterizing Smooth Homogeneous Manifolds}\label{AppConChaTheorems}
I should quickly rehearse two old definitions before making a new one. Firstly, let $H$ be a group with some group action, $\theta:H\times\mathcal{M}\to\mathcal{M}$, on a smooth manifold, $\mathcal{M}$. The stabilizer subgroup of $H$ at a point $p_0\in\mathcal{M}$ is the subgroup of $H$ whose group action on $\mathcal{M}$ maps $p_0$ to itself as $\theta(h,p_0)=p_0$. That is, the stabilizer subgroup of $H$ at $p_0$ is the subgroup of $H$ which maps $p_0$ onto itself. 

Secondly, a group $G_2$ is said to be a core-free subgroup of $G_1$ if and only if $G_2$ contains no normal subgroups of $G_1$ (besides the trivial group, $\{\openone\}$). Note that this implies that $G_2$ itself is not a normal subgroup of $G_1$ (unless, of course, $G_2$ is itself trivial). 

The following definition and theorem show how one can construct smooth homogeneous manifolds from a pair of Lie groups.
\begin{quote}
{\hypertarget{FCCompat}{\bf Definition: Faithful Construction Compatibility - }} Let $G_1$ and $G_2$ be two finite-dimensional Lie groups. These two Lie groups will be said to have \textit{faithful construction compatibility} if and only if $G_2$ is both a closed subgroup of $G_1$ as well as a core-free subgroup of $G_1$.
\end{quote}
For later reference, a fact regarding abelian groups should be noted. If $G_1$ is abelian, then all of its subgroups, $G_2$, are normal, $G_2\lhd G_1$. Hence if $G_1$ and $G_2$ are supposed to have faithful construction compatibility and $G_1$ is abelian then we must have $G_2=\{\openone\}$.
\begin{quote}
\hypertarget{HMConThm}{{\bf Homogeneous Manifold Construction Theorem}} - Let $G_1$ be a Lie group and $G_2\subset G_1$ a closed subgroup thereof. The quotient of these groups, $\mathcal{M}_\text{G}\coloneqq G_1/G_2$, is a smooth homogeneous manifold upon which $G_1$ acts smoothly and transitively via the following group action. Any $g\in G_1$ acts on any $[g_0]\in\mathcal{M}_\text{G}$ as $\bar{g}$ with,
\begin{align}
\bar{g}([g_0])=[g\,g_0]    ,
\end{align}
where $[g']\in\mathcal{M}_\text{G}$ is the $G_2$-coset in $G_1$ containing $g'\in G_1$. Together, these $\bar{g}$ maps form a Lie group of diffeomorphisms on $\mathcal{M}_\text{G}$, namely, $\overline{G}_1\subset\text{Diff}(\mathcal{M}_\text{G})$. These diffeomorphisms, $\overline{G}_1$, will be a faithful representation of $G_1$ (i.e., we will have $\overline{G}_1\cong G_1$) if and only if $G_1$ and $G_2$ have faithful construction compatibility.
\end{quote}
For proof see Sec.~\ref{AppHomoManConThm}. This construction theorem implies the well-known result that the quotient of a Lie group with one of its closed subgroups yields a smooth homogeneous manifold. The added content here is the condition under which the numerator group will be faithfully represented as diffeomorphisms on the resulting quotient space. 

The following theorem gives the converse result.
\begin{quote}
\hypertarget{HMChaThm}{{\bf Homogeneous Manifold Characterization Theorem}} - Let $\mathcal{M}$ be a smooth manifold. $\mathcal{M}$ is homogeneous if and only if there exists a finite-dimensional Lie group of its own diffeomorphisms, $H_\text{trans}\subset\text{Diff}(\mathcal{M})$, which acts transitively over $\mathcal{M}$. Let $H_\text{fix}$ be any stabilizer subgroup of $H_\text{trans}$. Any such pair of Lie groups, $H_\text{trans}$ and $H_\text{fix}$, will have faithful construction compatibility. Hence, by the 
Homogeneous Manifold Construction Theorem we can build a smooth manifold from them. The resulting manifold is diffeomorphic to the original,
\begin{align}\label{HomoRecon}
\mathcal{M}\cong \mathcal{M}_\text{H}\coloneqq H_\text{trans}/H_\text{fix}.
\end{align}
This reconstruction of $\mathcal{M}$ is robust in the sense that, up to diffeomorphism, it does not matter which $H_\text{trans}$ and $H_\text{fix}$ we use.
\end{quote}
For proof see Sec.~\ref{AppHomoManChaThm}. This characterization theorem implies the well-known result that every smooth homogeneous manifold, $\mathcal{M}$, can be reconstructed, up to diffeomorphism, as the quotient of two Lie groups. The added content here is that we can take these Lie groups to have faithful construction compatibility and, moreover, to be subgroups of $\mathcal{M}$'s own diffeomorphisms, $H_\text{fix}\subset H_\text{trans}\subset\text{Diff}(\mathcal{M})$. For a concrete demonstration of this reconstruction technique for the 2-sphere, $S^2\cong \text{SO}(3)/\text{SO}(2)$, see Sec.~\ref{AppSphereRecon}.

\subsection{Proof of the Homogeneous Manifold Construction Theorem}\label{AppHomoManConThm}
This section will prove the Homogeneous Manifold Construction Theorem. Given any finite-dimensional Lie group, $G_1$, and closed subgroup thereof,
$G_2\subset G_1$, one can build a homogeneous manifold as follows. Taking the quotient of these Lie groups yields the following quotient space,
\begin{align}\label{MGDef0}
\mathcal{M}_\text{G}\coloneqq G_1/G_2.
\end{align}
The equivalence classes under this quotient, $[g]=g\,G_2$, are $G_2$-cosets. Note that the quotient between these groups is, in general, not itself a Lie group. This is because the denominator, $G_2$, may not be a normal subgroup of the numerator, $G_1$. The quotient is, however, always a smooth manifold. In particular, this quotient space naturally adopts a smooth structure from the Lie group in the numerator, $G_1$, via its quotient with $G_2$. This follows from the quotient manifold theorem:\footnote{See Theorem 1 of \cite{Zikidis}}
\begin{quote}
{\hypertarget{QMT}{\bf Quotient Manifold Theorem - }} Suppose $G$ is a Lie group acting smoothly, freely, and properly on a smooth manifold $M$. Then the quotient space $M/G$ has a unique smooth structure with the property that the quotient map $\pi: M\to M/G$ is a smooth submersion. (N.b.: If $M$ is a Lie group and $G$ is a closed subgroup thereof, then $G$ acts smoothly, freely, and properly on $M$.)
\end{quote}
In order to apply this to our current situation we take the replacements $G=G_2$ and $M=G_1$ such that $M/G=\mathcal{M}_\text{G}$. The fact that $G_2$ acts smoothly, freely, and properly on $G_1$ follows from $G_2$ being a closed subgroup of $G_1$. Hence we can grant $\mathcal{M}_\text{G}$ the smooth structure which is uniquely picked out by this theorem.

In terms of diffeomorphisms on $\mathcal{M}_\text{G}$ this means the following: A generic transformation, \mbox{$d_\text{post}:\mathcal{M}_\text{G}\to \mathcal{M}_\text{G}$}, is smooth on $\mathcal{M}_\text{G}$ if and only if there exists a smooth transformation, \mbox{$d_\text{pre}:G_1\to G_1$}, which maps the equivalence classes onto each other in the same way,
\begin{align}
\pi\circ d_\text{pre}=d_\text{post}\circ\pi.
\end{align}
That is, some permutation of the equivalence classes, $[g]\mapsto d_\text{post}([g])$, will be smooth if and only if there is a way to implement it smoothly on $G_1$ prior to taking the quotient, i.e., to implement it as $[g]\mapsto [d_\text{pre}(g)]$.

This quotient space, $\mathcal{M}_\text{G}$, is not only a smooth manifold, but is also homogeneous with $G_1$ acting smoothly and transitively over it. Namely, the following group action,
\begin{align}\label{gbarDef}
&\theta:G_1\times\mathcal{M}_\text{G}\to\mathcal{M}_\text{G}\\
\nonumber
&\theta(g,[h])=\bar{g}([h])\coloneqq[g\,h].
\end{align}
acts smoothly and transitively over $\mathcal{M}_\text{G}$. This follows straightforwardly from the left-action of $G_1$ on itself being both smooth and transitive.

Given that this group action is smooth, the above defined $\bar{g}$ maps are all diffeomorphisms on $\mathcal{M}_\text{G}$. Collecting these together we have a Lie group $\bar{g}\in\overline{G_1}\subset\text{Diff}(\mathcal{M}_\text{G})$. In some sense, the Lie group $\overline{G_1}$ is a representation of $G_1$ as diffeomorphisms on $\mathcal{M}_\text{G}$. As I will now prove, this will be a faithful representation (i.e., $\overline{G_1}\cong G_1$) if and only $G_2$ is a core-free subgroup of $G_1$. (Recall that $G_2$ is a core-free subgroup of $G_1$ if and only if $G_2$ contains no non-trivial normal subgroups of $G_1$.)

The desired congruence result, $\overline{G_1}\cong G_1$, is equivalent to the map $g\mapsto\bar{g}$ having a trivial kernel. Namely, it is equivalent to the claim that $\bar{s}([g])=[g]$ implies $s=\openone$. But $\bar{s}([g])=[g]$ is equivalent to $[g^{-1}s g]=[e]$ which is itself equivalent to $g^{-1}s g\in G_2$. Hence our desired congruence result, $\overline{G_1}\cong G_1$, is equivalent to the following:
\begin{align}
\forall s \left( (\forall g \ \ g^{-1}s\,g\in G_2)\ \Longrightarrow \ s=\openone\right)
\end{align}
with $s$ and $g$ both ranging over $G_1$. Let $N_{G_1}(G_2)$ denote the set of all $s\in G_1$ satisfying $\forall g \ \ g^{-1}\,s\,g\in G_2$. Note that $N_{G_1}(G_2)\subset G_2$ (consider taking $g=\openone$). Note also that $N_{G_1}(G_2)\lhd G_1$ is a normal subgroup of $G_1$. In fact, $N_{G_1}(G_2)$ is the largest normal subgroup of $G_1$ which is contained within $G_2$. This is called the normal core of $G_2$ in $G_1$. Hence our desired congruence result, $\overline{G_1}\cong G_1$, is equivalent to $N_{G_1}(G_2)={\openone}$. This happens exactly when $G_2$ contains no non-trivial normal subgroups of $G_1$. That is, when $G_2$ is a core-free subgroup of $G_1$.

\subsection{Proof of the Homogeneous Manifold Characterization Theorem}\label{AppHomoManChaThm}
This section will prove the Homogeneous Manifold Characterization Theorem. By definition, a smooth manifold $\mathcal{M}$ is homogeneous if and only if there exists some finite-dimensional Lie group, $G$, which acts smoothly and transitively over it. Suppose that $G$ acts smoothly on $\mathcal{M}$ with action $\theta:G\times\mathcal{M}\to \mathcal{M}$. Using this action we can associate to each $g\in G$ a diffeomorphism $h\in\text{Diff}(\mathcal{M})$ with $h(p)=\theta(g,p)$. Collecting these diffeomorphisms together we have a finite-dimensional Lie group of diffeomorphisms, $H_\text{trans}\subset\text{Diff}(\mathcal{M})$. Note that $H_\text{trans}$ acts transitively on $\mathcal{M}$ (because $G$ does). Note also that $H_\text{trans}$ automatically acts smoothly on $\mathcal{M}$ since it is a group of diffeomorphisms on $\mathcal{M}$. Hence, $\mathcal{M}$ is homogeneous if and only if there exists some finite-dimensional Lie group of its diffeomorphisms, $H_\text{trans}\subset\text{Diff}(\mathcal{M})$, which acts transitively over $\mathcal{M}$.

I will now prove that $H_\text{trans}$ has faithful construction compatibility with any of its stabilizer subgroups $H_\text{fix}$ (say the stabilizer subgroup at $p_0$). To begin, let us prove that $H_\text{fix}$ is a closed subgroup of $H_\text{trans}$. Note that the function $f_{p_0}:H_\text{trans}\to\mathcal{M}$ given $f(h)=h(p_0)$ is continuous. Next note that $H_\text{fix}=f^{-1}(p_0)$. Since the singleton $
\{p_0\}$ is a closed set and $f$ is continuous, we have that $H_\text{fix}$ is a closed set.

Next, let us prove that $H_\text{fix}$ is a core-free subgroup of $H_\text{trans}$. As noted above, this claim is equivalent to the following:
\begin{align}
\forall s \left((\forall h \ \ h^{-1}s\,h\in H_\text{fix})\ \Longrightarrow \ s=\openone_\mathcal{M}\right)
\end{align}
with $s$ and $h$ both ranging over $H_\text{trans}$.
This equivalent claim will now be proved by contrapositive. Note that if $s\neq \openone_\mathcal{M}$ then there exists some \mbox{$q\in\mathcal{M}$} with \mbox{$s(q)\neq q$}. Since $H_\text{trans}$ acts transitively on $\mathcal{M}$ there is some $h\in H_\text{trans}$ which maps $p_0$ to $q$ as $q=h(p_0)$. For this $h$ we have \mbox{$(s\,h)(p_0)\neq q$} whereas $h(p_0)=q$. Thus we have \mbox{$(h^{-1}s\,h)(p_0)\neq p_0$} and therefore $h^{-1}s\,h\not\in H_\text{fix}$ as desired. This completes the proof that there is faithful construction compatibility between any transitively acting Lie group $H_\text{trans}\subset\text{Diff}(\mathcal{M})$ and any of its stabilizer subgroups, $H_\text{fix}$. 

Let us next use the above-proved construction theorem to build a homogeneous manifold from $H_\text{trans}$ and $H_\text{fix}$, namely $\mathcal{M}_\text{H}\coloneqq H_\text{trans}/H_\text{fix}$. According to the characterization theorem, this should be diffeomorphic to the original manifold. But how do we know? To establish that $\mathcal{M}_\text{H}$ and $\mathcal{M}$ are diffeomorphic we need to find a bijection between them which is smooth in both directions (i.e., a diffeomorphism). Before picking out such a bijection, let us rewrite the quotient space $\mathcal{M}_\text{H}\coloneqq H_\text{trans}/H_\text{fix}$ using the orbit stabilizer theorem: The $H_\text{fix}$-cosets in $H_\text{trans}$ are exactly the equivalence classes of the following equivalence relation,
\begin{align}
h_2\equiv_{p_0} h_1 \text{ iff } h_2(p_0)=h_1(p_0).
\end{align}
That is, $h_2$ and $h_1$ are equivalent under $\equiv_{p_0}$ if they map $p_0$ to the same place. Given that $h H_\text{fix}=[h]_{p_0}$, let us consider the map, 
\begin{align}\label{PDef}
P:\mathcal{M}_\text{H}&\to \mathcal{M}\\
\nonumber
[h]_{p_0}&\mapsto h(p_0).
\end{align}
This $P$ map sends each equivalence class, $[h]_{p_0}$, onto the point, $h(p_0)\in\mathcal{M}$, which all of its elements jointly map $p_0$ onto. Note that no two equivalence classes will send $p_0$ to the same place (by definition). Hence, $P$ is injective. Next note that $P$ is surjective because $H_\text{trans}$ acts transitively on $\mathcal{M}$ (i.e., for any $q$ there is some $h$ with $h(p_0)=q$). Hence, $P$ a bijection. Thus we have,
\begin{align}
\mathcal{M}\leftrightarrow \mathcal{M}_\text{H}\coloneqq H_\text{trans}/H_\text{fix}.
\end{align}
But is this bijection $P$ between $\mathcal{M}$ and $\mathcal{M}_\text{H}$ also a smooth map? It is, in fact, as can be seen from the following fundamental theorem of differential geometry:\footnote{See Theorem 3 of \cite{Zikidis}}
\begin{quote}
{\hypertarget{UST}{\bf Unique Smoothness Theorem:}} Let $G$, $Q$, and $M$ be smooth manifolds and \mbox{$\pi:G\to Q$} a surjective smooth submersion. If \mbox{$F:G\to M$} is a smooth map that is constant across $\pi$ (that is, $\pi(p) = \pi(q) \Longrightarrow  F(p)=F(q)$) then there exist a unique smooth map $R : Q \to M$ such that $R\circ\pi = F$
\end{quote}
In order to apply this to our current situation we take the replacements $G = H_\text{trans}$, $Q=\mathcal{M}_\text{H}$, $M=\mathcal{M}$, and \mbox{$F(h)=h(p_0)$}. Note that each of $G=H_\text{trans}$, $Q=\mathcal{M}_\text{H}$, and $M=\mathcal{M}$ are smooth manifolds by either definition or construction. The map $\pi:H_\text{trans}\to \mathcal{M}_\text{H}$ is a smooth submersion by construction via the quotient manifold theorem. The $F$ map is constant across $\pi$ by construction.
 
Under these replacements, this theorem tells us that there exists a unique smooth map, $R$, such that \mbox{$R([h]_{p_0})=h(p_0)$}. But this is exactly the $P$ map defined in Eq.~\eqref{PDef}. Thus we have that $P=R$ is a smooth map as well as a bijection. To complete our proof that $P$ is a diffeomorphism, we could next show that $P^{-1}$ is smooth. Alternatively, we could show that $P$ has a constant rank. This follows from $P$ being equivariant under $H_\text{trans}$'s joint action on $\mathcal{M}$ and $\mathcal{M}_\text{H}$,
\begin{align}
P([g\,h]_{p_0})=g\,h\,p_0 = g\,P([h]_{p_0}).
\end{align}
Hence, we have that $P$ is a diffeomorphism and therefore that $\mathcal{M}\cong \mathcal{M}_\text{H}=H_\text{trans}/H_\text{fix}$ as smooth manifolds. Indeed, we have $\mathcal{M}=P(H_\text{trans}/H_\text{fix})$.

\subsection{Reconstructing the Sphere from its  Diffeomorphisms}\label{AppSphereRecon}
\begin{figure}[p]
\centering
\includegraphics[width=0.95\textwidth]{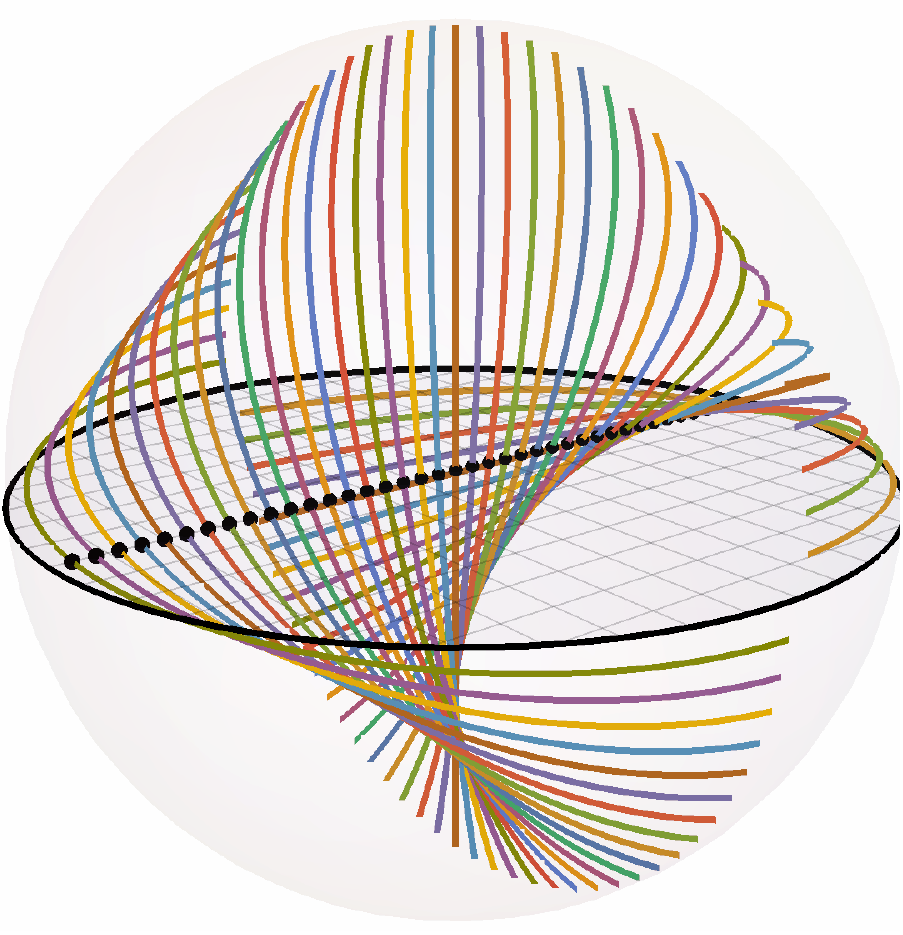}
\caption{The Lie group $\text{SO}(3)\cong\text{RP}^3$ is depicted here as a smooth manifold: Namely, as a ball with radius $\pi$ in $\mathbb{R}^3$ with every pair of antipodal points on its surface identified. Concretely, each point $r\in\mathbb{R}^3$ in this figure represents a rotation of $\theta=\vert r\vert$ radians about the $\hat{r}$ axis (where $\hat{r}$ is the unit vector pointing at $r$). We can recover the sphere $S^2$ from $\text{SO}(3)$ by taking its quotient with respect to a certain equivalence relation, $\equiv_{p_0}$. Two rotations are equivalent according to $\equiv_{p_0}$ if they map the north-pole, $p_0=n$, to the same place. The colored lines shown are some of the equivalence classes of $\text{SO}(3)$ under $\equiv_{p_0}$. Note that the equator itself is an equivalence class. For every equivalence class except this one, we can pick out a unique representative on the $x,y$-plane (see the black points). Taking these points as our representatives the quotient space becomes a disk with its boundary compactified to a single point. Namely, this quotient of $\text{SO}(3)$ by $\equiv_{p_0}$ yield the smooth manifold $\text{SO}(3)/\equiv_{p_0}\cong S^2$ as desired.}\label{FigSO3}
\end{figure}

To explicitly demonstrate how these theorems work, I will now use them to reconstruct the 2-sphere, $\mathcal{M}\cong S^2$. The reader is invited to think of $\mathcal{M}\cong S^2$ as a time-less version of the spacetime manifold, $\mathcal{M}_\text{old}\cong \mathbb{R}\times S^2$, from the Quartic Klein-Gordon theory introduced in Sec.~\ref{SecSurInt}. Correspondingly, the relevant Lie group of diffeomorphisms for the following discussion, \mbox{$H_\text{trans}\cong\text{SO}(3)$}, is a time-less version of the $H_\text{trans}\cong(\mathbb{R},+)\times \text{SO}(3)$ group discussed at the end of Sec.~\ref{SecSurInt}. The removal of the time dimension from this example is merely for pedagogical reasons; Removing time allows us to visualize the following discussion in terms of a nice three-dimensional picture, namely Fig.~\ref{FigSO3}.

In order to reconstruct $\mathcal{M}\cong S^2$ from $H_\text{trans}$ we must first identify $H_\text{fix}$, the stabilizer subgroup of $H_\text{trans}$. Recall that \mbox{$H_\text{trans}\cong\text{SO}(3)$} is the group of rigid rotations of $\mathcal{M}\cong S^2$ in some coordinate system. If we take $p_0=n$ to be the north pole, then we have that $H_\text{fix}\cong\text{SO}(2)$ is the group rotations about the z-axis. Plugging $H_\text{trans}\cong\text{SO}(3)$ and $H_\text{fix}\cong\text{SO}(2)$ into Eq.~\eqref{HomoRecon} we have, 
\begin{align}\label{SO3SO2}
\mathcal{M}_\text{H}\coloneqq H_\text{trans}/H_\text{fix}\cong SO(3)/SO(2).
\end{align}
According to the construction theorem, this quotient should be a smooth manifold. Moreover, according to the characterization theorem, it should be diffeomorphic to our original manifold, $\mathcal{M}_\text{H}\cong \mathcal{M}\cong S^2$. Let us now check this, paying careful attention to where the quotient manifold gets its smooth structure from. See Fig.~\ref{FigSO3} for an illustrated guide to how this quotient works.

In Eq.~\eqref{SO3SO2} we are supposed to regard the Lie group in the numerator, $H_\text{trans}\cong\text{SO}(3)$, as a smooth manifold: But what smooth manifold is $\text{SO}(3)$? To see this, consider first its Lie algebra, $\mathfrak{so}(3)$, and note that it is three-dimensional. The group exponential from $\mathfrak{so}(3)$ to $\text{SO}(3)$ is surjective such that each element of $\text{SO}(3)$ is represented (possibly multiple times) in the image of $\mathfrak{so}(3)$ under the group exponential. As I will now discuss, by identifying these redundancies we can see $\text{SO}(3)$ as a certain quotient space of $\mathbb{R}^3$. 

Fig.~\ref{FigSO3} illustrates exactly how $\mathfrak{so}(3)$ redundantly represents $\text{SO}(3)$. In Fig.~\ref{FigSO3} a point, $r\in\mathbb{R}^3$, represents a rotation of $\theta=\vert r\vert$ radians about the $\hat{r}$ axis (where $\hat{r}$ is the unit vector pointing at $r$). In this way, each rotation in $\text{SO}(3)$ can represented within the region $\vert r\vert\leq \pi$ with the following redundancy: $\pi\,\hat{r}$ is identical to $-\pi\,\hat{r}$ for every $\hat{r}$. That is, a half-turn about any axis $\hat{r}$ is the same as a half-turn in the reverse direction.

For example, rotations of the sphere about the $x$-axis correspond to points on the $x$-axis in Fig.~\ref{FigSO3} (e.g., the black points). Rotating about the $x$-axis through an angle just greater than $\pi$ is equivalent to rotating in the reverse direction by an angle just less than $\pi$. Hence, the antipodal points on the $x$-axis in Fig.~\ref{FigSO3} represent the same rotation and so should be identified. This holds not just the $x$-axis but for every axis. Thus, as a smooth manifold $\text{SO}(3)$ is diffeomorphic to the unit ball in $\mathbb{R}^3$ with every pair of antipodal points on its surface identified. That is, as a smooth manifold we have $\text{SO}(3)\cong \text{RP}^3$, the real-projective space in three dimensions.

Note that $\text{RP}^3$ is not diffeomorphic to the manifold, $\mathcal{M}=S^2$, which we are hoping to recover. Indeed, according to Eq.~\eqref{SO3SO2} the desired manifold is a quotient of this manifold by $H_\text{fix}\cong\text{SO}(2)$. The colored lines in Fig.~\ref{FigSO3} show some of these $H_\text{fix}$-cosets.\footnote{Note that only a subset of these $H_\text{fix}$-cosets (or equivalently, $\equiv_{p_0}$-equivalence classes) are shown in Fig.~\ref{FigSO3}. Namely, only equivalence classes which contain rotations about the $x$-axis with incrementally larger values of $\theta\in[-\pi,\pi]$ are shown.}

At this point, it will be helpful to use the orbit-stabilizer theorem to rewrite the quotient manifold, $\mathcal{M}_\text{H}\coloneqq H_\text{trans}/H_\text{fix}$, in terms of an equivalence relation. This theorem tells us that $H_\text{fix}$-cosets in $H_\text{trans}$ are identical to the equivalence classes of the following equivalence relation over $H_\text{trans}$: 
\begin{align}
h_2\equiv_{p_0} h_1 \text{ iff } h_2(p_0)=h_1(p_0),
\end{align}
where $p_0=n$ is the point at which $H_\text{fix}$ is $H_\text{trans}$'s stabilizer subgroup. In other words, two diffeomorphisms, $h_1,h_2\in H_\text{trans}$ are equivalent under $\equiv_{p_0}$ if they map $p_0=n$ to the same place. The orbit-stabilizer theorem says that we have $h H_\text{fix}=[h]_{p_0}$, an exact matching between $H_\text{fix}$-cosets and $\equiv_{p_0}$ equivalence classes.  Thus, the reconstructed manifold can be rewritten as,
\begin{align}
\mathcal{M}_\text{H}
= H_\text{trans}/\equiv_{p_0}.
\end{align}
We can understand Fig.~\ref{FigSO3} in terms of this equivalence relation as follows: the colored lines in Fig.~\ref{FigSO3} show which rotations in $H_\text{trans}\cong\text{SO}(3)$ map the north-pole, $p_0=n$ to the same place.

Let us quickly discuss two examples of equivalence classes. First, note that all rotations about the z-axis map the north-pole to the same place (i.e., they leave it fixed). Hence, every z-axis rotation is in the same $\equiv_{p_0}$ equivalence class. Rotations about the z-axis are represented in Fig.~\ref{FigSO3} by points on the z-axis.  Hence, these points are connected by a vertical line in Fig.~\ref{FigSO3}. Second, note that all rotations about points on the equator through an angle of $\theta=\pi$ map the north-pole to the same place (i.e., to the south pole). Hence, all of these rotations are in the same $\equiv_{p_0}$ equivalence class. In Fig.~\ref{FigSO3}, these rotations are represented by points on the sphere's equator. Hence, the equator is connected by a circular line in Fig.~\ref{FigSO3}. Besides these two equivalence classes, several others are also shown in Fig.~\ref{FigSO3} which smoothly transition between them.

But how can we see the sphere $S^2$ arising from this collection of $\equiv_{p_0}$-equivalence class? For every equivalence class (except for the equatorial one) we can pick as its representative element its unique member on the $x,y$-plane (e.g., the black points in Fig.~\ref{FigSO3}). The exception to this claim is the equivalence class of rotations which map the north-pole to the south-pole (i.e., the black line around the equator). Just like every other equivalence class, we must treat the equator in Fig.~\ref{FigSO3} like a single point in the quotient space. Thus, we can take the $x,y$-plane of Fig.~\ref{FigSO3} to be our quotient space with one slight modification: We must treat the equator as being compactified into a single point in the quotient space. Compactifying the boundary of a disk yields the manifold $\mathcal{M}_\text{H}\cong S^2$ as desired.